\documentclass[aps,prl,twocolumn,superscriptaddress,reprint]{revtex4-1}

\usepackage{blindtext}
\usepackage{centernot}
\usepackage{graphicx}
\usepackage{amsmath,bbold}
\usepackage{times}
\usepackage{amssymb}
\usepackage{mathrsfs}
\usepackage{chemarr}
\usepackage{color}
\usepackage{url}
\usepackage{version}
\usepackage[hidelinks]{hyperref}
\usepackage{mwe,tikz}
\usepackage[percent]{overpic}
\usepackage{bm}
\usepackage[export]{adjustbox}
\definecolor{linkcolor}{rgb}{0,0,0.6} 
\usepackage{ stmaryrd }

\newcommand{\bJ}{{\bf J}}

\newcommand{\br}{{\bf r}}

\newcommand{\bu}{{\bf u}}
\newcommand\grad{\nabla}

\newcommand{\dd}{\mathrm{d}}

\usepackage{lipsum}

\usetikzlibrary{patterns}

\newcommand\ra{ {\rm a} }

\begin{document}

\title{Lamellar to micellar phases and beyond: when tactic active systems admit free-energy functionals}

\author{J. O'Byrne}
\affiliation{Universit\'e de Paris, Laboratoire Mati\`ere et Syst\`emes Complexes (MSC), UMR 7057 CNRS, F-75205 Paris, France}

\author{J. Tailleur}
\affiliation{Universit\'e de Paris, Laboratoire Mati\`ere et Syst\`emes Complexes (MSC), UMR 7057 CNRS, F-75205 Paris, France}

\date{\today}

\begin{abstract}
  We consider microscopic models of active particles whose velocities,
  rotational diffusivities, and tumbling rates depend on the gradient
  of a local field, which is either externally imposed or depends on
  all particle positions.  Despite the fundamental differences between
  active and passive dynamics at the microscopic scale, we show that a
  large class of such tactic active systems admit fluctuating
  hydrodynamics equivalent to those of interacting Brownian colloids
  in equilibrium.  We exploit this mapping to show how taxis may lead
  to the lamellar and micellar phases observed for soft repulsive
  colloids. In the context of chemotaxis, we show how the competition
  between chemoattractant and chemorepellent may lead to a bona-fide
  equilibrium liquid-gas phase separation in which a loss of
  thermodynamic stability of the fluid signals the onset of a
  chemotactic collapse.
\end{abstract}

\maketitle

Over the past ten years, the development of a wealth of synthetic
active systems has paved the way for engineering active
materials~\cite{howse2007self,palacci2010sedimentation,thutupalli2011swarming,bricard2013emergence,palacci2013living,nishiguchi2015mesoscopic,yan2016reconfiguring,bechinger2016active,moerman2017solute}. Unlike
in equilibrium, however, there is no guiding principle for the
self-assembly of active systems, due to the lack of a generic
expression for their steady states. A natural way forward, which has
been heavily investigated
recently~\cite{tailleur2008statistical,speck2014effective,takatori2014swim,ginot2015nonequilibrium,marconi2016pressure,rodenburg2017van,solon2018generalized,flenner2020active},
is then to determine to which extent the knowledge we have garnered
in---and close to---equilibrium remains relevant to active matter.
A first positive answer is provided by the wealth of works on
effective temperatures in active
systems~\cite{loi2008effective,Wang2011JCP,loi2011effective,morozov2010motor,szamel2014self,solon2015active}
which show that an effective fluctuation-dissipation relation may,
under some conditions, survive activity. Then, similarities between
active and passive dynamics have also been detected at the level of
collective
behaviours~\cite{chavanis2004,golestanian2012collective,takatori2015towards,paliwal2018chemical}. For
instance, motility-induced phase separation (MIPS), which happens at
large persistence lengths and times, relies on a feedback between
activity and kinetic hindrance without counterparts in thermal
equilibrium. Nevertheless, it shares several features, at a
coarse-grained level, with an equilibrium liquid-gas phase
separation~\cite{tailleur2008statistical,thompson2011lattice,speck2014effective,takatori2014swim,ginot2015nonequilibrium,bi2016motility,grafke2017spatiotemporal}. A
natural question is then how general this meso-to-large-scale
similarity between active and passive dynamics is, and under which
conditions, if any, the microscopic drive out of equilibrium of active
particles disappears upon coarse-graining~\cite{nardini2017entropy}.

Among the physical phenomena that control active dynamics, taxis plays
an important role in a large range of situations~\footnote{Taxis
  describes the biased motion of an entity in response to an external
  signal}. It is widespread in the biological world, from the
chemotaxis of run-and-tumble bacteria~\cite{berg2008coli} to the
phototaxis of \textit{algae}~\cite{polin2009chlamydomonas} through the
durotaxis of cells~\cite{sunyer2016collective}.
{Taxis} is however not limited to living systems: self-propelled
colloids often rely on the presence of chemicals in their environment
that not only power their self-propulsion~\cite{howse2007self} but
also bias their dynamics~\cite{soto2014self,pohl2014dynamic}. {Taxis}
is known to lead to rich, system-specific behaviours, that occur both
at the microscopic and collective
levels~\cite{woodward1995spatio,brenner1998physical,chatterjee2011chemotaxis,saragosti2011directional,saha2014clusters,soto2014self,pohl2014dynamic,agudo2019active,nasouri2020exact}.
{While some of these phenomena are  nonequilibrium in
  nature~\cite{goldstein1996traveling,saragosti2011directional,seyrich2019traveling,agudo2019active,mahdisoltani2019controlled},
  the aforementioned connections between active systems and
  equilibrium physics raises the question as to whether microscopic
  tactic dynamics may also lead to emerging behaviours described by
  coarse-grained equilibrium theories.}

In this Letter, we answer this question for a broad class of active
particles, which we refer to as `tactic active particles' (TAPs),
whose propelling speeds or orientational dynamics are biased by the
gradients of a field $c(\br,t)$. Using a diffusive scaling of time and
space, we construct their fluctuating hydrodynamics and determine
under which conditions the latter satisfy detailed balance, whence
admitting a free-energy functional. This allows us to reveal a one-to-one
correspondance, at this coarse-grained level, between tactic active
particles and interacting Brownian colloids. Despite the fundamental
differences between these two systems at the microscopic scales, a
mapping may thus exist between the large-scale dynamics and
steady-state statistics of their density fields. We then exploit this
result to report the existence of the micellar and lamellar phases
usually observed for softly repulsive Brownian colloids. In the
context of chemotaxis, we further show how the competition between
chemoattractant and chemorepellent can be rationalized using the
physics of Brownian colloids undergoing a liquid-gas phase
separation. A simple criterion is then proposed to predict the onset
of a chemotactic collapse, which amounts to a loss of thermodynamic
stability of the fluid phase. Finally, we show that externally imposed
chemical fields $c(\br)$ map onto external potentials in equilibrium.

We consider $N$ self-propelled particles whose
positions evolve as
\begin{equation}
  \dot{\mathbf{r}}_i=v_p \mathbf{u}_i + \sqrt{2 D_t} \boldsymbol{\eta}_i\;,
  \label{EDS}
\end{equation}
where $\bu_i$ is a unit vector indicating the orientation of particle
$i$, $v_p$ is its self-propulsion speed, and
$\{\boldsymbol{\eta}_i\}$ forms a set of $N$ Gaussian white noises
whose spatial components satisfy
$\left \langle \eta_i^{a}(t) \eta_j^{b}(t')\right \rangle =
\delta^{ab} \delta_{ij} \delta(t-t')$. The particles' orientations
evolve through rotational diffusion (in $d>1$ dimensions) and instantaneous
tumbles. To model {taxis}, we consider two possible couplings between
the field $c$ and the active dynamics. First, the speed of a
particle may depend on its orientation with respect to $\grad c$,
through a linear coupling
\begin{equation}\label{eq:chemospeed}
v_p = v_0 - v_1 \mathbf{u}_i\cdot \nabla c
\end{equation}
Alternatively, {taxis} may stem from the anisotropy of the
orientational dynamics. We model the latter as direction-dependent
tumbling rate $\alpha$ and rotational diffusivity
$\Gamma$~\footnote{The rotational diffusion is understood as an
  It\=o-Langevin process.}:
\begin{equation}\label{eq:chemodirection}
\alpha = \alpha_0 + \alpha_1 \mathbf{u}_i\cdot \nabla c \hspace{0.2cm}
\text{and} \hspace{0.2cm} \Gamma = \Gamma_0 + \Gamma_1\mathbf{u}_i\cdot \nabla c
\end{equation}
Positive values of $v_1$, $\alpha_1$, and
$\Gamma_1$ drive the particles towards lower values of
$c$.

Experimentally, Eq.~\eqref{eq:chemospeed} can be implemented using
feedback loops~\cite{bauerle2018self,lavergne2019group} when the speed
of self-propelled particles can be controlled by light---a class that
comprises both Janus
colloids~\cite{palacci2013living,buttinoni2013dynamical,bauerle2018self}
and
bacteria~\cite{vizsnyiczai2017light,frangipane2018dynamic,arlt2018painting,arlt2019dynamics}. The
field $c(\br,\{\br_i\})$ can then be an arbitrary function of all
particle positions. Equation~\eqref{eq:chemodirection} is also a
standard model for the chemotaxis of run-and-tumble
bacteria~\cite{schnitzer1993theory,berg2008coli,chatterjee2011chemotaxis,seyrich2019traveling}. The
field $c(\br)$ then models chemorepellent ($\alpha_1>0$) or
chemoattractant ($\alpha_1<0$), which can be either produced by the
bacteria or imposed externally.
To cover both cases, we consider the many-body dynamics in which
$c(\br_i,[\rho])$ is both a function of $\br_i$ and a functional of
the particle density. In the context of diffusing fields, this amounts
to integrating out the dynamics of the chemotactic fields, following,
e.g.,~\cite{soto2014self,pohl2014dynamic}.

\textit{Fluctuating mesoscopic description.}  While much insight can
be gained from deterministic hydrodynamic descriptions of active
systems~\cite{bertin2006boltzmann,liebchen2017phoretic,kourbane2018exact,seyrich2019traveling},
measuring their irreversibility by, say, computing entropy production
requires working at the fluctuating level.  At scales much larger than
the persistence length of the particles, we expect the active random
walks to lead to a diffusive behaviour. We thus rescale space and time
as $(x,t) \to (x/L,t/L^2)$, where $L$ is the linear size of the
system. To lighten the notations, we focus for now on rotational
diffusion. The $N$-body probability density
$\psi(\{\br_i,\bu_i\},t)$ then evolves as
\begin{eqnarray}\label{eq:MastEq}
\partial_t\psi &=&-L \sum_{i=1}^N\nabla _{\br_i}\cdot\Big[\Big(v_0-\frac {v_1}L \bu_i \cdot \grad_{\br_i} c\Big)  \bu_i\psi- \frac{D_t}L\nabla_{\br_i}\psi\Big]\notag\\
& & +L^2\sum_i\Delta_{\bu_i} \Big[\big(\Gamma_0+\frac{\Gamma_1}L \bu_i \cdot \grad_{\br_i} c\Big) \psi\Big] 
\end{eqnarray}
Integrating over all orientational degrees of freedom leads to a
conservation equation
$\partial_t \phi(\{\br_i\},t) = - \sum_i \grad_{\br_i} \cdot
\bJ_i(\{\br_i\},t)$, where
$\bJ_i\equiv -D_t\grad_{\br_i} \phi+\int \Pi_j \dd\bu_j \left(v_0L
  -v_1\bu_i\cdot \grad_{\br_i}c\right)\bu_i \psi$ is the probability
current along direction $\br_i$ and $\phi$ is the marginal in space of
$\psi$. To compute the diffusive limit of ${\bJ}_i$, the
dynamics~\eqref{eq:MastEq} can be projected onto its successive
$N$-body harmonics, leading to a hierarchy of equations~\cite{supp}.
Under the asumption of a lack of long-range correlations between the particle
orientations, a closure is obtained by noticing that all moments
beyond $\phi(\{\br_i\},t)$ are fast fields and decay as $1/L$ (or
faster) in the large system-size limit, leading to:
\begin{eqnarray}\label{eq:diffcurrent}
  \mathbf{J}_i &\simeq& -\Big(D_t+\frac{v_0^2}{d(d-1)\Gamma_0}\Big)\nabla_{\br_i}\phi - \phi \Big[\frac{\grad_{\br_i} v_0^2}{2d(d-1) \Gamma_0}\notag \\&& 
  +\Big(\frac{v_0 \Gamma_1+v_1\Gamma_0}{d \Gamma_0 }\Big) \nabla_{\br_i}c\Big]
\end{eqnarray}
At this stage, we have constructed a coarse-grained diffusive
description of dynamics~\eqref{EDS}-\eqref{eq:chemodirection}. We
now restore, for full generality, the possibility of tumbles and
take $v_i$, $\alpha_i$, and $\Gamma_i$ constant to focus on the
consequences of taxis. Using stochastic calculus, one then obtains a
corresponding fluctuating hydrodynamics for the density field
$\rho(\br)=\sum_{i=1}^N \delta(\br-\br_i)$:
\begin{eqnarray}
  \partial_t \rho &=&- \nabla\cdot [J_D+\sqrt{2D\rho}\Lambda] \label{hydrorho}\\ J_D
  &=& -D \rho \grad\Big[\log \rho  + \Big( v_1 + v_0\frac{\alpha_1+(d-1)\Gamma_1}{\alpha_0 + (d-1)\Gamma_0} \Big) \frac{c}{dD}\Big]
\label{hydroJ}
\end{eqnarray}
where $\Lambda(\br,t)$ is a Gaussian noise field of zero mean and
correlations $\langle \Lambda(\br,t) \Lambda(\br',t')\rangle=
\delta(t-t') \delta(\br-\br')$, $J_D$ is the deterministic part of the
current, and $D$ is a large-scale diffusivity:
\begin{equation}\label{eq:D}
D = D_t + \frac{v_0^2}{d(\alpha_0+(d-1)\Gamma_0)}\;.
\end{equation}
Equations~\eqref{hydrorho} and~\eqref{hydroJ} describe the stochastic
dynamics of the density field of TAPs at scales much larger than their
persistence lengths and times, and can now be used to study their
emerging collective behaviours.


\textit{Effective free-energy functionals.}  Inspection of
Eq.~\eqref{hydroJ} shows that the deterministic part of the current
can be written as $J_D=-D \rho \grad \mu$, where $\mu$ plays the role
of a non-equilibrium chemical
potential~\cite{chaikin1995principles}. An interesting outcome of the
diffusive scaling is that the noise field and the mobility appearing
in $J_D$ satisfy a generalized Stokes-Einstein relation. The
dynamics~\eqref{hydrorho} then satisfies detailed balance whenever one
can find a free energy ${\cal F}[\rho]$ whose functional derivative is
given by $\mu$. 
\begin{figure*}[t]
  \begin{center}
    \begin{tikzpicture}
      \def\x{3.5}
      \def\xx{.15}
      \def\y{-3.2}
      \draw (-2,0) node[rotate=90] {\large Active};
      
      \draw (0,0) node {\includegraphics[totalheight=3.25cm]{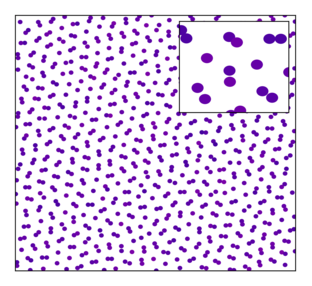}};
      \draw (\x,0) node{\includegraphics[width=.2\textwidth]{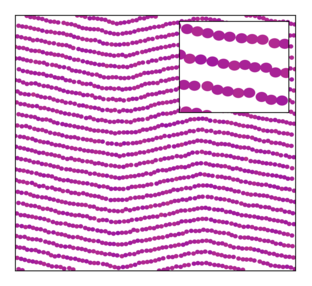}};
      \draw (2*\x,0) node{\includegraphics[width=.2\textwidth]{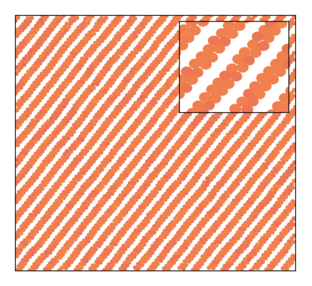}};
      \draw (3*\x+\xx,0) node{\includegraphics[width=.22\textwidth]{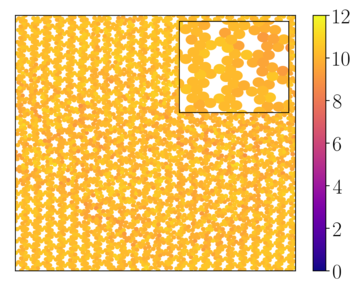}};

      \draw (3*\x+\xx+2.2,1.4) node {\small $\rho$};

      \draw (-2,\y) node[rotate=90] {\large Passive};
      \draw (0,\y) node{\includegraphics[width=.2\textwidth]{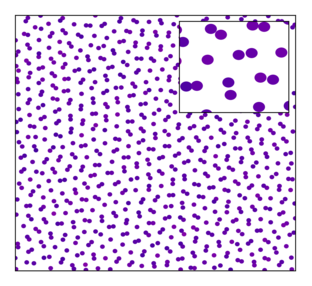}};
      \draw (\x,\y) node{\includegraphics[width=.2\textwidth]{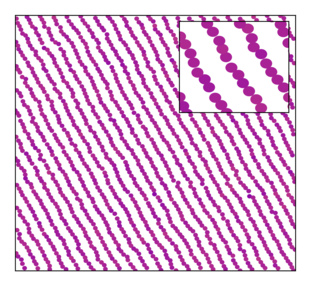}};
      \draw (2*\x,\y) node{\includegraphics[width=.2\textwidth]{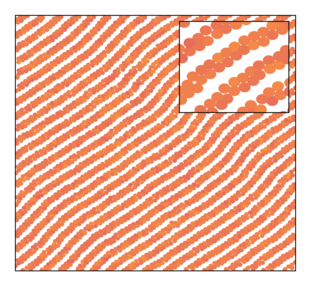}};
      \draw (3*\x+\xx,\y) node{\includegraphics[width=.22\textwidth]{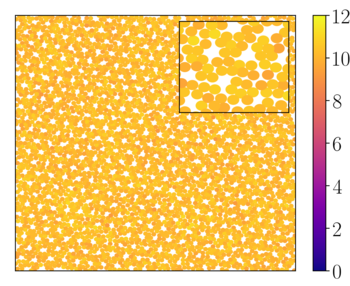}};
      \draw (3*\x+\xx+2.2,\y+1.4) node {\small $\rho$};
    \end{tikzpicture}
  \end{center}
  \caption{\label{pattern} Simulations of RTP
    dynamics~\eqref{EDS}-\eqref{eq:chemospeed} (top) and of
    the equilibrium dynamics~\eqref{EDS:collloids} (bottom), under the
    conditions of the mapping~\eqref{eq:mapping} with $K(\br)$ defined
    by \eqref{eq:twoshoulder}. Color encodes the local
    density. Parameters: $v_0 =1$, $v_1 = 0.2$, $\alpha_0 = 50$,
    $D_t = 0$, $\sigma_0 = 0.3$, $\sigma_1 = 1$, $\varepsilon = 5$,
    $E = 10$, $dt=10^{-3}$. Snapshots taken at $t=20\,000$ in a system
    of size $20\times20$. From left to right, $\rho_0\equiv N/L^2$
    is equal to 2 , 4 , 8 and 10.}
\end{figure*}
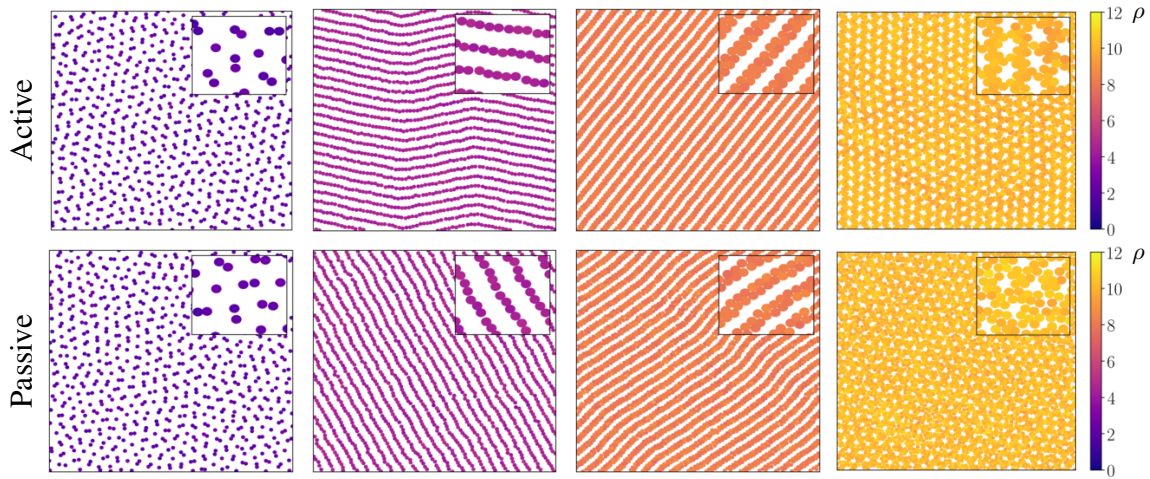
The solution of such an inverse variational
problem~\cite{anderson1989variational} can be obtained by generalizing
the Schwarz condition of integrability to functional integration. In
practice, we introduce
\begin{equation}\label{eq:defD}
  \mathcal{D}(\mathbf{r},\mathbf{r'})=
  \frac{\delta \mu([\rho],\mathbf{r})}{\delta \rho(\mathbf{r'})} -
  \frac{\delta \mu([\rho],\mathbf{r'})}{\delta \rho(\mathbf{r})}\;,
\end{equation}
which is such that $\mu(\br,[\rho])$ is a functional derivative iif,
for any two test functions $f,\,g$~\footnote{As usual, there are
  topological requirement for the conditions to be sufficient~\cite{anderson1989variational}},
\begin{equation}
  \int \mathcal{D}(\mathbf{r},\mathbf{r'})f(\mathbf{r})g(\mathbf{r'})\mathrm{d}\mathbf{r}\mathrm{d}\mathbf{r'}=0\;.
  \label{Schwarz}
\end{equation}
{Note that Eq.~\eqref{Schwarz} means that $\mathcal{D}$ vanishes as a
distribution, which is not always straightforward to read in its
expression. For instance, in one dimension, $\mu=\partial_x^k \rho$
leads to
$\mathcal{D}(x,x')=(\partial_x^k-\partial_{x'}^k)\delta(x-x')$; $\mu$
admits a functional integral iif $k$ is even, which can be checked
easily using~\eqref{Schwarz}.}

Let us now consider a field $c$ given by:
\begin{equation}\label{convol}
    c([\rho],\br) = \int K(\br,\br_1,...,\br_{p-1}) \rho(\br_1)...\rho(\br_{p-1})\mathrm{d}\br_1...\mathrm{d}\br_{p-1}
\end{equation}
Equation~\eqref{Schwarz} is satisfied whenever the kernel $K$ is
invariant under any permutation of its variables. The fluctuating
hydrodynamics~\eqref{hydrorho}-\eqref{hydroJ} then becomes:
\begin{equation}
\partial_t \rho = \nabla\cdot \Big[ D\rho \nabla \frac{\delta {\cal F}}{\delta \rho} +\sqrt{2D\rho}\Lambda \Big]
\label{hydroEq}
\end{equation}
where the effective free-energy functional ${\cal F}$ is given by:
\begin{equation}\label{hydrofree}
\begin{aligned}
    {\cal F}[\rho] = &\int \mathrm{d}\br \rho(\br) \log \rho(\br)  + \frac{v_0}{d D} \Big[ \frac{v_1}{v_0} + \frac{\alpha_1+(d-1)\Gamma_1}{\alpha_0 + (d-1)\Gamma_0} \Big] \times \\ &  \frac{1}{p!}\int  K(\br_1,...,\br_p)\rho(\br_1)...\rho(\br_p) \mathrm{d}\br_1...\mathrm{d}\br_p
\end{aligned}
\end{equation}
Importantly, Eqs.~\eqref{hydroEq} and~\eqref{hydrofree} also describe the
fluctuating hydrodynamics of $N$ Brownian colloids interacting via a
$p$-body potential. More precisely, consider $N$ particles undergoing
the equilibrium Langevin dynamics
\begin{equation}
  \gamma\dot{\br}_i = -\sum_{j_1<...<j_{p-1}} \nabla_{\br_i} V(\br_{i},\br_{j_1},...,\br_{j_{p-1}}) +\sqrt{2\gamma T}\boldsymbol{\eta}\;,
  \label{EDS:collloids}
\end{equation}
where we have introduced a temperature $T$, a damping $\gamma$ and a
$p$-body potential $V$. Equations~\eqref{hydroEq} and~\eqref{hydrofree}
describe the fluctuating hydrodynamics of this system,
upon identifying:
\begin{eqnarray}\label{eq:mapping}
\gamma^{-1} T &\Leftrightarrow&  D \\  \gamma^{-1} V(\br_1,...,\br_p) &\Leftrightarrow&   \Big[ \frac{v_1}{d} + \frac{v_0}{d}\frac{\alpha_1+(d-1)\Gamma_1}{\alpha_0 + (d-1)\Gamma_0} \Big]K(\br_1,...,\br_p)\;.\notag
\end{eqnarray}
The mapping~\eqref{eq:mapping} establishes a macroscopic connection
between the non-equilibrium
dynamics~\eqref{EDS}-\eqref{eq:chemodirection} and the equilibrium
dynamics~\eqref{EDS:collloids}, which strongly differs at the
microscopic scale. In particular, the roles played by $K$ and $V$ in
these microscopic dynamics are of very different natures, despite
their similar role in the macroscopic dynamics of $\rho$. This differs
from approaches in which the chemotaxis of cells is already modelled
at the microscopic level by Brownian dynamics in which the field $c$
plays the role of a potential~\cite{chavanis2010stochastic}. The
macroscopic equivalence described there with equilibrium dynamics
directly stems from a microscopic one. Note that, while a similarity
between active and passive collective behaviours has been reported in
the case of
MIPS~\cite{tailleur2008statistical,thompson2011lattice,speck2014effective,takatori2014swim,ginot2015nonequilibrium,bi2016motility,grafke2017spatiotemporal},
the mapping~\eqref{eq:mapping} suggests a much broader relationship
between passive and active systems, which we now explore.

\textit{Micellar, lamellar and crystalline phases.} The recent
development of active systems in which the particle velocities can be
controlled individually using light and feedback
loops~\cite{bauerle2018self,lavergne2019group} makes the realization
of dynamics~\eqref{EDS}-\eqref{eq:chemospeed} within experimental
reach. To explore and rationalize the physics these systems can access, we consider a
simple case in which the field $c$ is obtained using a convolution
kernel $K(\br,\br_1)=K(\br-\br_1)$ with two typical lengthscales:
\begin{equation} \label{eq:twoshoulder}
K(\mathbf{r})= A \,{\rm e}^{-\frac{\sigma_0^2}{\sigma_0^2-r^2}}\Theta(\sigma_0 - |\mathbf r | )+ \epsilon \, {\rm e}^{-\frac{\sigma_1^2}{\sigma_1^2-r^2}}\Theta(\sigma_1 - | \mathbf r | )\;,
\end{equation}
where $\Theta(u)$ is the Heaviside function and we consider
$A > \epsilon$ and $\sigma_1>\sigma_0$. We report in the top row of
Fig.~\ref{pattern} simulations of the active
dynamics~\eqref{EDS}-\eqref{eq:chemospeed}.  As the density increases,
the system undergoes a series of phase transitions: a disordered gas
(not shown) is first replaced by a crystal of `micelles', each
comprising an increasing number of particles as the density increases
($1^{\rm st}$ column), until a laning transition occurs. Further
increasing the density increases the local number of particles forming
the lanes ($2^{\rm nd}\&3^{\rm rd}$ columns). At higher densities,
inverted crystals develop, in which voids and particles have exchanged
their previous roles ($4^{\rm th}$ column). The underlying physics can
be rationalized thanks to our equilibrium mapping: simulations of the
passive dynamics~\eqref{EDS:collloids} for parameters
satisfying~\eqref{eq:mapping} indeed perfectly match those of TAPs
(Fig.~\ref{pattern}, bottom row). In the passive picture, the
kernel~\eqref{eq:twoshoulder} corresponds to a pairwise potentials
with a repulsive central core and a softer repulsive shoulder. Using
equilibrium Monte Carlo simulations, similar potentials have been
shown to lead to a variety of exotic
phases~\cite{malescio2003stripe,glaser2007soft,lowen2011applications},
which are thus accessible to TAPS.

\textit{Chemotaxis: competition between attraction and repulsion.}
While fields $c(\br, [\rho])$ given by~\eqref{convol} can be engineered in the lab,
they can also be found in nature. In biological systems, tactic
interactions are often mediated by diffusing molecules, which
naturally lead to a linear coupling between $\rho$ and
$c$. Interactions mediated by fluctuating membranes or interfaces
would lead to non-linearities which offer an interesting (and
challenging) problem left for future works. Inspired by bacterial
chemotaxis, we consider a model system in which active particles
interact through the production of a chemoattractant and a
chemorepellent. Introducing their concentration fields $c_{\rm a}$ and
$c_{\rm r}$, we write the tumbling rate of particle $i$ as
$\alpha=\alpha_0 - \alpha_1^{\ra} \bu_i \cdot \nabla c_\ra +
\alpha_1^{\rm r} \bu_i \cdot \nabla c_{\rm r}$. Taking
$\alpha_1^{\rm r}>0$ and $\alpha_1^{\rm a}>0$ then bias the random
walk of particle $i$ towards high-$c_{\rm a}$ regions and away from
high-$c_{\rm r}$ regions.

\begin{figure}
    \centering
\includegraphics[width=.30\columnwidth]{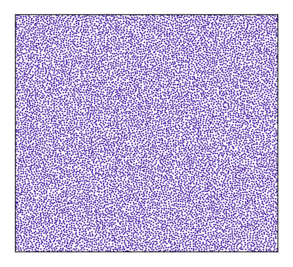}
\includegraphics[width=.30\columnwidth]{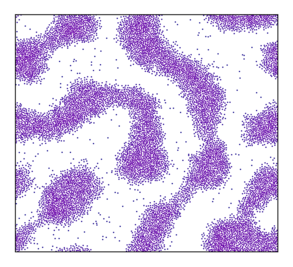}
\includegraphics[width=.365\columnwidth]{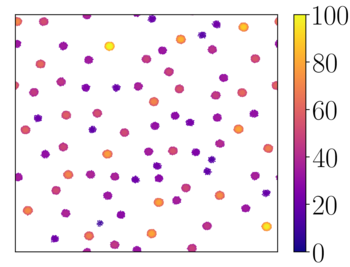}

\includegraphics[width=.30\columnwidth]{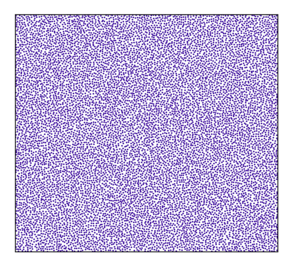}
\includegraphics[width=.30\columnwidth]{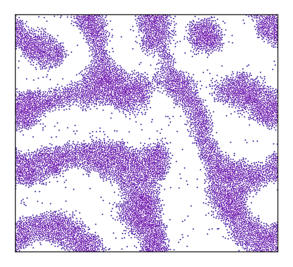}
\includegraphics[width=.365\columnwidth]{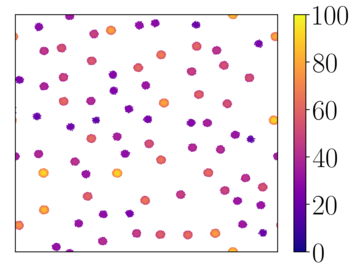}
\caption{Simulations of the active dynamics (1)-(3), top, and the equilibrium dynamics (14), bottom, using the pair-potential~\eqref{eq:pairpotentialchemo}, under the conditions of the mapping~\eqref{eq:mapping}. Color encodes the local
  density. From left to right $\ell_{\rm r} = 1/6 ,1/8, 1/10$. A chemotactic
  collapse is predicted whenever $\bar V<0$, i.e. below
  $\ell_{\rm r}=1/8.2$. Parameters: $L=40$, $\rho_0=7$,
  $a_{\rm a}/2\pi \nu_{\rm a}=0.05$, $\ell_{\rm a}=0.25$, $a_{\rm r}/2\pi \nu_{\rm r}=0.2$,
  $\alpha_0=100$, $\alpha_1^{\rm r}=\alpha_1^{\rm a}=200$, $v_0=1$,
  $T^{-1}=200$, $\gamma =1$,  $\mathrm{d}t =
  10^{-3}$. The snapshots were taken after a time
  $t=10^{4}$.} \label{fig:collapse}

\end{figure}

After their production by the particles, at rates $a_{\rm a,r}$, the signaling molecules diffuse with diffusivity $\nu_{\rm a,r}$ and are degraded at rates $\lambda_{\rm a,r}$, leading to the dynamics
\begin{equation}
\partial_t c_{\rm n}(\br) = \nu_{\rm n} \Delta c_{\rm n}(\br)-\lambda_{\rm n} c_{\rm n}+a_{\rm n}\sum_i \delta(\br-\br_i)
\label{eq:KS}
\end{equation}
where $n \in \{{\rm a}, {\rm r}\}$. 
A standard, fast variable treatment on $c_{\rm n}$ then leads to the screened Poisson equation 
\begin{equation}
\Big(\Delta - \frac{\lambda_{\rm n}}{\nu_{\rm n}}\Big)c_{\rm n} = -\frac{a_{\rm n}}{\nu_{\rm n}}\rho\;,
\label{ScreenedPoisson}
\end{equation}
which can be solved as $c_{\rm n}=\frac{a_{\rm n}}{\nu_{\rm n}} G_{\rm n}\ast \rho$, where
$G_{\rm n}$ is the Green function of Eq.~\eqref{ScreenedPoisson}. The system
thus satisfies the mapping condition~\eqref{convol} with
$p=2$. Introducing the screening lengths
$\ell_{\rm n}\equiv \sqrt{\nu_{\rm n}/\lambda_{\rm n}}$, the Green functions are given by
$G_{\rm n}(\br)=\ell_{\rm n} e^{-|\br \vert/\ell_{\rm n}}/2$ in 1D,
$G_{\rm n}(\br)=K_0(|\br|/\ell_{\rm n})/2\pi$ in 2D, where $K_0$ is the $0^{th}$
order modified Bessel function of the second kind, and
$K(\br)=e^{-|\br|/\ell_{\rm n}}/4\pi |\br|$ in 3D. The
mapping~\eqref{eq:mapping} then shows this system of active particles
to be equivalent, at the fluctuating hydrodynamic level, to passive
Brownian particles interacting via a pair potential
\begin{equation}\label{eq:pairpotentialchemo}
  V (\br)= \frac{\gamma
  v_0}{d\alpha_0}\Big[\alpha_1^{\rm r}\frac{a_{\rm r}}{\nu_{\rm r}}G_{\rm r}(\br) -
\alpha_1^{\rm a}\frac{a_{\rm a}}{\nu_{\rm a}}G_{\rm a}(\br)\Big]\;.
\end{equation}
The superposition of
chemoattractant and chemorepellent thus directly maps onto an
equilibrium problem with attractive and repulsive interactions. Let us
consider the case in which $\alpha_1^{\rm r}>\alpha_1^{\rm a}$ and
$\ell_{\rm r} \leq \ell_{\rm a}$. The physics of this chemotactic system is now
mapped onto the well-known problem of repulsive hard-core interactions
with attractive tails. The system is purely repulsive for
$\ell_{\rm a}=\ell_{\rm r}$, hence leading to a gas phase. As $\ell_{\rm r}$ decreases,
the attractive tail develops, allowing for a liquid-gas
coexistence. For even shorter $\ell_{\rm r}$, the liquid phase becomes
thermodynamically unstable leading to a collapse of the system when
$\bar V\equiv \int \dd{\br}
V(\br)<0$~\cite{ruelle1999statistical}. Figure~\ref{fig:collapse}
compares simulations of the passive dynamics~\eqref{EDS:collloids} and
the active dynamics~\eqref{EDS}-\eqref{eq:chemodirection}, using the
2D Green functions of Eq.~\eqref{ScreenedPoisson}, as $\ell_{\rm r}$ is
varied. Once again, despite the fundamental differences between their
microscopic dynamics and interactions, the large-scale physics of
these two systems are hardly distinguishable. 

\textit{External fields.} Finally, while we have focused so
far on collective effects, nothing prevents an additional taxis towards an
externally controled field $w(\br)$. Our models of
taxis~\eqref{eq:chemospeed} and~\eqref{eq:chemodirection} are easily
generalized to this case, using for instance:
\begin{equation}
\left\{\begin{matrix}
v_p &=& v_0 - v_1 \mathbf{u}_i\cdot \nabla c -v_2 \mathbf{u}_i\cdot \nabla w\\ 
\alpha &=& \alpha_0 + \alpha_1 \mathbf{u}_i\cdot \nabla c + a_2 \mathbf{u}_i\cdot \nabla w \\ 
\Gamma &=& \Gamma_0 + \Gamma_1\mathbf{u}_i\cdot \nabla c +\Gamma_2 \mathbf{u}_i\cdot \nabla w
\end{matrix}\right.
\end{equation}
where the field $c$ still accounts for interactions between the self-propelled particles given by Eq.~\eqref{convol}. The equilibrium mapping still holds, with a free energy now given by:
\begin{eqnarray*}
\tilde {\cal F}[\rho] = {\cal F}[\rho] +  \frac{v_0}{d D} \Big[ \frac{v_2}{v_0} + \frac{\alpha_2+(d-1)\Gamma_2}{\alpha_0 + (d-1)\Gamma_0} \Big]\int {\rm d} \br w(\br)\rho(\br)
\end{eqnarray*}
where ${\cal F}$ remains given by~Eq.~\eqref{hydrofree}. The external
field $w(\br)$ is thus equivalent to an external potential for
Brownian colloids. Much like gravity, it can be used to localize a
dense phase in a phase-separated system.

\textit{Discussion.} The mapping between TAPS and Brownian colloids
presented in this article offers an interesting route to control
synthetic active matter, for instance to prepare a desired initial
condition in a light-controlled active
system~\cite{bauerle2018self,lavergne2019group}. It also offers a
qualitative insight into the large-scale behaviours of tactic active
particles, which, as for MIPS, should hold beyond
the sole systems that will obey exactly the
dynamics~\eqref{EDS}-\eqref{eq:chemodirection}.

The derivation of Eq.~\eqref{eq:diffcurrent}, detailed in
SI~\cite{supp}, works directly at the interacting $N$-body level. This
differs from standard treatment of chemotaxis that rely on an explicit
dynamics of the field $c$, assumed to be slow, to work at an
essentially non-interacting level~\cite{othmer2000diffusion}. Our
treatment relies on a lack of long-range orientational order and
precludes, in particular, large-scale collective motion. Whether the
latter may emerge in a setting as simple as the one described
by~\eqref{EDS}-~\eqref{eq:chemodirection} is an open challenging
question, stimulated in particular by the works on bacterial
travelling
waves~\cite{saragosti2011directional,seyrich2019traveling}. Furthermore,
our derivation assumes that the field gradients remain finite, which
has to be checked self-consistently. In particular, it certainly does
not hold in the late-stage dynamics of the chemotactic collapse
reported in Fig.~\ref{fig:collapse}, which only highlights the
remarkable agreement between passive and active dynamics in this case.

In this Letter, we have focused on active systems whose large-scale
dynamics obey detailed balance, but the interest of our mapping is not
limited to these cases. Studying the linear response to fields
violating~\eqref{Schwarz} or~\eqref{eq:mapping} is a natural next
step. Near-equilibrium studies of active matter models have indeed
been shown to capture many interesting properties of active
systems~\cite{marconi2016velocity,fodor2016far,wittmann2017effective1,wittmann2017effective2}.
Furthermore, the approach to the inverse variational problem described in
Eq.~\eqref{eq:defD}-\eqref{Schwarz} can easily be generalized to more complex
situations, for instance involving several species of active
particles~\cite{wittkowski2017nonequilibrium,agudo2019active,curatolo2019cooperative,kolb2020active}.

Finally, we have focused on the consequences that the
mapping~\eqref{eq:mapping} bears on the steady-state distributions of
tactic active systems. The one-to-one correspondence is, however,
established at a dynamical level which paves the way to a study of the
transition states of tactic active systems, using methods like
transition path sampling~\cite{dellago1998transition} or the string
method~\cite{weinan2002string}.

\bibliographystyle{apsrev4-1}
\bibliography{biblio}

\begin{thebibliography}{78}%
\makeatletter
\providecommand \@ifxundefined [1]{%
 \@ifx{#1\undefined}
}%
\providecommand \@ifnum [1]{%
 \ifnum #1\expandafter \@firstoftwo
 \else \expandafter \@secondoftwo
 \fi
}%
\providecommand \@ifx [1]{%
 \ifx #1\expandafter \@firstoftwo
 \else \expandafter \@secondoftwo
 \fi
}%
\providecommand \natexlab [1]{#1}%
\providecommand \enquote  [1]{``#1''}%
\providecommand \bibnamefont  [1]{#1}%
\providecommand \bibfnamefont [1]{#1}%
\providecommand \citenamefont [1]{#1}%
\providecommand \href@noop [0]{\@secondoftwo}%
\providecommand \href [0]{\begingroup \@sanitize@url \@href}%
\providecommand \@href[1]{\@@startlink{#1}\@@href}%
\providecommand \@@href[1]{\endgroup#1\@@endlink}%
\providecommand \@sanitize@url [0]{\catcode `\\12\catcode `\$12\catcode
  `\&12\catcode `\#12\catcode `\^12\catcode `\_12\catcode `\%12\relax}%
\providecommand \@@startlink[1]{}%
\providecommand \@@endlink[0]{}%
\providecommand \url  [0]{\begingroup\@sanitize@url \@url }%
\providecommand \@url [1]{\endgroup\@href {#1}{\urlprefix }}%
\providecommand \urlprefix  [0]{URL }%
\providecommand \Eprint [0]{\href }%
\providecommand \doibase [0]{http://dx.doi.org/}%
\providecommand \selectlanguage [0]{\@gobble}%
\providecommand \bibinfo  [0]{\@secondoftwo}%
\providecommand \bibfield  [0]{\@secondoftwo}%
\providecommand \translation [1]{[#1]}%
\providecommand \BibitemOpen [0]{}%
\providecommand \bibitemStop [0]{}%
\providecommand \bibitemNoStop [0]{.\EOS\space}%
\providecommand \EOS [0]{\spacefactor3000\relax}%
\providecommand \BibitemShut  [1]{\csname bibitem#1\endcsname}%
\let\auto@bib@innerbib\@empty
\bibitem [{\citenamefont {Howse}\ \emph {et~al.}(2007)\citenamefont {Howse},
  \citenamefont {Jones}, \citenamefont {Ryan}, \citenamefont {Gough},
  \citenamefont {Vafabakhsh},\ and\ \citenamefont
  {Golestanian}}]{howse2007self}%
  \BibitemOpen
  \bibfield  {author} {\bibinfo {author} {\bibfnamefont {J.~R.}\ \bibnamefont
  {Howse}}, \bibinfo {author} {\bibfnamefont {R.~A.}\ \bibnamefont {Jones}},
  \bibinfo {author} {\bibfnamefont {A.~J.}\ \bibnamefont {Ryan}}, \bibinfo
  {author} {\bibfnamefont {T.}~\bibnamefont {Gough}}, \bibinfo {author}
  {\bibfnamefont {R.}~\bibnamefont {Vafabakhsh}}, \ and\ \bibinfo {author}
  {\bibfnamefont {R.}~\bibnamefont {Golestanian}},\ }\href@noop {} {\bibfield
  {journal} {\bibinfo  {journal} {Physical Review Letters}\ }\textbf {\bibinfo
  {volume} {99}},\ \bibinfo {pages} {048102} (\bibinfo {year}
  {2007})}\BibitemShut {NoStop}%
\bibitem [{\citenamefont {Palacci}\ \emph {et~al.}(2010)\citenamefont
  {Palacci}, \citenamefont {Cottin-Bizonne}, \citenamefont {Ybert},\ and\
  \citenamefont {Bocquet}}]{palacci2010sedimentation}%
  \BibitemOpen
  \bibfield  {author} {\bibinfo {author} {\bibfnamefont {J.}~\bibnamefont
  {Palacci}}, \bibinfo {author} {\bibfnamefont {C.}~\bibnamefont
  {Cottin-Bizonne}}, \bibinfo {author} {\bibfnamefont {C.}~\bibnamefont
  {Ybert}}, \ and\ \bibinfo {author} {\bibfnamefont {L.}~\bibnamefont
  {Bocquet}},\ }\href@noop {} {\bibfield  {journal} {\bibinfo  {journal}
  {Physical Review Letters}\ }\textbf {\bibinfo {volume} {105}},\ \bibinfo
  {pages} {088304} (\bibinfo {year} {2010})}\BibitemShut {NoStop}%
\bibitem [{\citenamefont {Thutupalli}\ \emph {et~al.}(2011)\citenamefont
  {Thutupalli}, \citenamefont {Seemann},\ and\ \citenamefont
  {Herminghaus}}]{thutupalli2011swarming}%
  \BibitemOpen
  \bibfield  {author} {\bibinfo {author} {\bibfnamefont {S.}~\bibnamefont
  {Thutupalli}}, \bibinfo {author} {\bibfnamefont {R.}~\bibnamefont {Seemann}},
  \ and\ \bibinfo {author} {\bibfnamefont {S.}~\bibnamefont {Herminghaus}},\
  }\href@noop {} {\bibfield  {journal} {\bibinfo  {journal} {New Journal of
  Physics}\ }\textbf {\bibinfo {volume} {13}},\ \bibinfo {pages} {073021}
  (\bibinfo {year} {2011})}\BibitemShut {NoStop}%
\bibitem [{\citenamefont {Bricard}\ \emph {et~al.}(2013)\citenamefont
  {Bricard}, \citenamefont {Caussin}, \citenamefont {Desreumaux}, \citenamefont
  {Dauchot},\ and\ \citenamefont {Bartolo}}]{bricard2013emergence}%
  \BibitemOpen
  \bibfield  {author} {\bibinfo {author} {\bibfnamefont {A.}~\bibnamefont
  {Bricard}}, \bibinfo {author} {\bibfnamefont {J.-B.}\ \bibnamefont
  {Caussin}}, \bibinfo {author} {\bibfnamefont {N.}~\bibnamefont {Desreumaux}},
  \bibinfo {author} {\bibfnamefont {O.}~\bibnamefont {Dauchot}}, \ and\
  \bibinfo {author} {\bibfnamefont {D.}~\bibnamefont {Bartolo}},\ }\href@noop
  {} {\bibfield  {journal} {\bibinfo  {journal} {Nature}\ }\textbf {\bibinfo
  {volume} {503}},\ \bibinfo {pages} {95} (\bibinfo {year} {2013})}\BibitemShut
  {NoStop}%
\bibitem [{\citenamefont {Palacci}\ \emph {et~al.}(2013)\citenamefont
  {Palacci}, \citenamefont {Sacanna}, \citenamefont {Steinberg}, \citenamefont
  {Pine},\ and\ \citenamefont {Chaikin}}]{palacci2013living}%
  \BibitemOpen
  \bibfield  {author} {\bibinfo {author} {\bibfnamefont {J.}~\bibnamefont
  {Palacci}}, \bibinfo {author} {\bibfnamefont {S.}~\bibnamefont {Sacanna}},
  \bibinfo {author} {\bibfnamefont {A.~P.}\ \bibnamefont {Steinberg}}, \bibinfo
  {author} {\bibfnamefont {D.~J.}\ \bibnamefont {Pine}}, \ and\ \bibinfo
  {author} {\bibfnamefont {P.~M.}\ \bibnamefont {Chaikin}},\ }\href@noop {}
  {\bibfield  {journal} {\bibinfo  {journal} {Science}\ }\textbf {\bibinfo
  {volume} {339}},\ \bibinfo {pages} {936} (\bibinfo {year}
  {2013})}\BibitemShut {NoStop}%
\bibitem [{\citenamefont {Nishiguchi}\ and\ \citenamefont
  {Sano}(2015)}]{nishiguchi2015mesoscopic}%
  \BibitemOpen
  \bibfield  {author} {\bibinfo {author} {\bibfnamefont {D.}~\bibnamefont
  {Nishiguchi}}\ and\ \bibinfo {author} {\bibfnamefont {M.}~\bibnamefont
  {Sano}},\ }\href@noop {} {\bibfield  {journal} {\bibinfo  {journal} {Physical
  Review E}\ }\textbf {\bibinfo {volume} {92}},\ \bibinfo {pages} {052309}
  (\bibinfo {year} {2015})}\BibitemShut {NoStop}%
\bibitem [{\citenamefont {Yan}\ \emph {et~al.}(2016)\citenamefont {Yan},
  \citenamefont {Han}, \citenamefont {Zhang}, \citenamefont {Xu}, \citenamefont
  {Luijten},\ and\ \citenamefont {Granick}}]{yan2016reconfiguring}%
  \BibitemOpen
  \bibfield  {author} {\bibinfo {author} {\bibfnamefont {J.}~\bibnamefont
  {Yan}}, \bibinfo {author} {\bibfnamefont {M.}~\bibnamefont {Han}}, \bibinfo
  {author} {\bibfnamefont {J.}~\bibnamefont {Zhang}}, \bibinfo {author}
  {\bibfnamefont {C.}~\bibnamefont {Xu}}, \bibinfo {author} {\bibfnamefont
  {E.}~\bibnamefont {Luijten}}, \ and\ \bibinfo {author} {\bibfnamefont
  {S.}~\bibnamefont {Granick}},\ }\href@noop {} {\bibfield  {journal} {\bibinfo
   {journal} {Nature materials}\ }\textbf {\bibinfo {volume} {15}},\ \bibinfo
  {pages} {1095} (\bibinfo {year} {2016})}\BibitemShut {NoStop}%
\bibitem [{\citenamefont {Bechinger}\ \emph {et~al.}(2016)\citenamefont
  {Bechinger}, \citenamefont {Di~Leonardo}, \citenamefont {L{\"o}wen},
  \citenamefont {Reichhardt}, \citenamefont {Volpe},\ and\ \citenamefont
  {Volpe}}]{bechinger2016active}%
  \BibitemOpen
  \bibfield  {author} {\bibinfo {author} {\bibfnamefont {C.}~\bibnamefont
  {Bechinger}}, \bibinfo {author} {\bibfnamefont {R.}~\bibnamefont
  {Di~Leonardo}}, \bibinfo {author} {\bibfnamefont {H.}~\bibnamefont
  {L{\"o}wen}}, \bibinfo {author} {\bibfnamefont {C.}~\bibnamefont
  {Reichhardt}}, \bibinfo {author} {\bibfnamefont {G.}~\bibnamefont {Volpe}}, \
  and\ \bibinfo {author} {\bibfnamefont {G.}~\bibnamefont {Volpe}},\
  }\href@noop {} {\bibfield  {journal} {\bibinfo  {journal} {Reviews of Modern
  Physics}\ }\textbf {\bibinfo {volume} {88}},\ \bibinfo {pages} {045006}
  (\bibinfo {year} {2016})}\BibitemShut {NoStop}%
\bibitem [{\citenamefont {Moerman}\ \emph {et~al.}(2017)\citenamefont
  {Moerman}, \citenamefont {Moyses}, \citenamefont {Van Der~Wee}, \citenamefont
  {Grier}, \citenamefont {Van~Blaaderen}, \citenamefont {Kegel}, \citenamefont
  {Groenewold},\ and\ \citenamefont {Brujic}}]{moerman2017solute}%
  \BibitemOpen
  \bibfield  {author} {\bibinfo {author} {\bibfnamefont {P.~G.}\ \bibnamefont
  {Moerman}}, \bibinfo {author} {\bibfnamefont {H.~W.}\ \bibnamefont {Moyses}},
  \bibinfo {author} {\bibfnamefont {E.~B.}\ \bibnamefont {Van Der~Wee}},
  \bibinfo {author} {\bibfnamefont {D.~G.}\ \bibnamefont {Grier}}, \bibinfo
  {author} {\bibfnamefont {A.}~\bibnamefont {Van~Blaaderen}}, \bibinfo {author}
  {\bibfnamefont {W.~K.}\ \bibnamefont {Kegel}}, \bibinfo {author}
  {\bibfnamefont {J.}~\bibnamefont {Groenewold}}, \ and\ \bibinfo {author}
  {\bibfnamefont {J.}~\bibnamefont {Brujic}},\ }\href@noop {} {\bibfield
  {journal} {\bibinfo  {journal} {Physical Review E}\ }\textbf {\bibinfo
  {volume} {96}},\ \bibinfo {pages} {032607} (\bibinfo {year}
  {2017})}\BibitemShut {NoStop}%
\bibitem [{\citenamefont {Tailleur}\ and\ \citenamefont
  {Cates}(2008)}]{tailleur2008statistical}%
  \BibitemOpen
  \bibfield  {author} {\bibinfo {author} {\bibfnamefont {J.}~\bibnamefont
  {Tailleur}}\ and\ \bibinfo {author} {\bibfnamefont {M.}~\bibnamefont
  {Cates}},\ }\href@noop {} {\bibfield  {journal} {\bibinfo  {journal}
  {Physical Review Letters}\ }\textbf {\bibinfo {volume} {100}},\ \bibinfo
  {pages} {218103} (\bibinfo {year} {2008})}\BibitemShut {NoStop}%
\bibitem [{\citenamefont {Speck}\ \emph {et~al.}(2014)\citenamefont {Speck},
  \citenamefont {Bialk{\'e}}, \citenamefont {Menzel},\ and\ \citenamefont
  {L{\"o}wen}}]{speck2014effective}%
  \BibitemOpen
  \bibfield  {author} {\bibinfo {author} {\bibfnamefont {T.}~\bibnamefont
  {Speck}}, \bibinfo {author} {\bibfnamefont {J.}~\bibnamefont {Bialk{\'e}}},
  \bibinfo {author} {\bibfnamefont {A.~M.}\ \bibnamefont {Menzel}}, \ and\
  \bibinfo {author} {\bibfnamefont {H.}~\bibnamefont {L{\"o}wen}},\ }\href@noop
  {} {\bibfield  {journal} {\bibinfo  {journal} {Physical Review Letters}\
  }\textbf {\bibinfo {volume} {112}},\ \bibinfo {pages} {218304} (\bibinfo
  {year} {2014})}\BibitemShut {NoStop}%
\bibitem [{\citenamefont {Takatori}\ \emph {et~al.}(2014)\citenamefont
  {Takatori}, \citenamefont {Yan},\ and\ \citenamefont
  {Brady}}]{takatori2014swim}%
  \BibitemOpen
  \bibfield  {author} {\bibinfo {author} {\bibfnamefont {S.~C.}\ \bibnamefont
  {Takatori}}, \bibinfo {author} {\bibfnamefont {W.}~\bibnamefont {Yan}}, \
  and\ \bibinfo {author} {\bibfnamefont {J.~F.}\ \bibnamefont {Brady}},\
  }\href@noop {} {\bibfield  {journal} {\bibinfo  {journal} {Physical Review
  Letters}\ }\textbf {\bibinfo {volume} {113}},\ \bibinfo {pages} {028103}
  (\bibinfo {year} {2014})}\BibitemShut {NoStop}%
\bibitem [{\citenamefont {Ginot}\ \emph {et~al.}(2015)\citenamefont {Ginot},
  \citenamefont {Theurkauff}, \citenamefont {Levis}, \citenamefont {Ybert},
  \citenamefont {Bocquet}, \citenamefont {Berthier},\ and\ \citenamefont
  {Cottin-Bizonne}}]{ginot2015nonequilibrium}%
  \BibitemOpen
  \bibfield  {author} {\bibinfo {author} {\bibfnamefont {F.}~\bibnamefont
  {Ginot}}, \bibinfo {author} {\bibfnamefont {I.}~\bibnamefont {Theurkauff}},
  \bibinfo {author} {\bibfnamefont {D.}~\bibnamefont {Levis}}, \bibinfo
  {author} {\bibfnamefont {C.}~\bibnamefont {Ybert}}, \bibinfo {author}
  {\bibfnamefont {L.}~\bibnamefont {Bocquet}}, \bibinfo {author} {\bibfnamefont
  {L.}~\bibnamefont {Berthier}}, \ and\ \bibinfo {author} {\bibfnamefont
  {C.}~\bibnamefont {Cottin-Bizonne}},\ }\href@noop {} {\bibfield  {journal}
  {\bibinfo  {journal} {Physical Review X}\ }\textbf {\bibinfo {volume} {5}},\
  \bibinfo {pages} {011004} (\bibinfo {year} {2015})}\BibitemShut {NoStop}%
\bibitem [{\citenamefont {Marconi}\ \emph
  {et~al.}(2016{\natexlab{a}})\citenamefont {Marconi}, \citenamefont {Maggi},\
  and\ \citenamefont {Melchionna}}]{marconi2016pressure}%
  \BibitemOpen
  \bibfield  {author} {\bibinfo {author} {\bibfnamefont {U.~M.~B.}\
  \bibnamefont {Marconi}}, \bibinfo {author} {\bibfnamefont {C.}~\bibnamefont
  {Maggi}}, \ and\ \bibinfo {author} {\bibfnamefont {S.}~\bibnamefont
  {Melchionna}},\ }\href@noop {} {\bibfield  {journal} {\bibinfo  {journal}
  {Soft Matter}\ }\textbf {\bibinfo {volume} {12}},\ \bibinfo {pages} {5727}
  (\bibinfo {year} {2016}{\natexlab{a}})}\BibitemShut {NoStop}%
\bibitem [{\citenamefont {Rodenburg}\ \emph {et~al.}(2017)\citenamefont
  {Rodenburg}, \citenamefont {Dijkstra},\ and\ \citenamefont {van
  Roij}}]{rodenburg2017van}%
  \BibitemOpen
  \bibfield  {author} {\bibinfo {author} {\bibfnamefont {J.}~\bibnamefont
  {Rodenburg}}, \bibinfo {author} {\bibfnamefont {M.}~\bibnamefont {Dijkstra}},
  \ and\ \bibinfo {author} {\bibfnamefont {R.}~\bibnamefont {van Roij}},\
  }\href@noop {} {\bibfield  {journal} {\bibinfo  {journal} {Soft matter}\
  }\textbf {\bibinfo {volume} {13}},\ \bibinfo {pages} {8957} (\bibinfo {year}
  {2017})}\BibitemShut {NoStop}%
\bibitem [{\citenamefont {Solon}\ \emph {et~al.}(2018)\citenamefont {Solon},
  \citenamefont {Stenhammar}, \citenamefont {Cates}, \citenamefont {Kafri},\
  and\ \citenamefont {Tailleur}}]{solon2018generalized}%
  \BibitemOpen
  \bibfield  {author} {\bibinfo {author} {\bibfnamefont {A.~P.}\ \bibnamefont
  {Solon}}, \bibinfo {author} {\bibfnamefont {J.}~\bibnamefont {Stenhammar}},
  \bibinfo {author} {\bibfnamefont {M.~E.}\ \bibnamefont {Cates}}, \bibinfo
  {author} {\bibfnamefont {Y.}~\bibnamefont {Kafri}}, \ and\ \bibinfo {author}
  {\bibfnamefont {J.}~\bibnamefont {Tailleur}},\ }\href@noop {} {\bibfield
  {journal} {\bibinfo  {journal} {Physical Review E}\ }\textbf {\bibinfo
  {volume} {97}},\ \bibinfo {pages} {020602} (\bibinfo {year}
  {2018})}\BibitemShut {NoStop}%
\bibitem [{\citenamefont {Flenner}\ and\ \citenamefont
  {Szamel}(2020)}]{flenner2020active}%
  \BibitemOpen
  \bibfield  {author} {\bibinfo {author} {\bibfnamefont {E.}~\bibnamefont
  {Flenner}}\ and\ \bibinfo {author} {\bibfnamefont {G.}~\bibnamefont
  {Szamel}},\ }\href@noop {} {\bibfield  {journal} {\bibinfo  {journal} {arXiv
  preprint arXiv:2004.11925}\ } (\bibinfo {year} {2020})}\BibitemShut {NoStop}%
\bibitem [{\citenamefont {Loi}\ \emph {et~al.}(2008)\citenamefont {Loi},
  \citenamefont {Mossa},\ and\ \citenamefont {Cugliandolo}}]{loi2008effective}%
  \BibitemOpen
  \bibfield  {author} {\bibinfo {author} {\bibfnamefont {D.}~\bibnamefont
  {Loi}}, \bibinfo {author} {\bibfnamefont {S.}~\bibnamefont {Mossa}}, \ and\
  \bibinfo {author} {\bibfnamefont {L.~F.}\ \bibnamefont {Cugliandolo}},\
  }\href@noop {} {\bibfield  {journal} {\bibinfo  {journal} {Physical Review
  E}\ }\textbf {\bibinfo {volume} {77}},\ \bibinfo {pages} {051111} (\bibinfo
  {year} {2008})}\BibitemShut {NoStop}%
\bibitem [{\citenamefont {Wang}\ and\ \citenamefont
  {Wolynes}(2011)}]{Wang2011JCP}%
  \BibitemOpen
  \bibfield  {author} {\bibinfo {author} {\bibfnamefont {S.}~\bibnamefont
  {Wang}}\ and\ \bibinfo {author} {\bibfnamefont {P.~G.}\ \bibnamefont
  {Wolynes}},\ }\href {\doibase 10.1063/1.3624753} {\bibfield  {journal}
  {\bibinfo  {journal} {The Journal of Chemical Physics}\ }\textbf {\bibinfo
  {volume} {135}},\ \bibinfo {pages} {051101} (\bibinfo {year}
  {2011})}\BibitemShut {NoStop}%
\bibitem [{\citenamefont {Loi}\ \emph {et~al.}(2011)\citenamefont {Loi},
  \citenamefont {Mossa},\ and\ \citenamefont {Cugliandolo}}]{loi2011effective}%
  \BibitemOpen
  \bibfield  {author} {\bibinfo {author} {\bibfnamefont {D.}~\bibnamefont
  {Loi}}, \bibinfo {author} {\bibfnamefont {S.}~\bibnamefont {Mossa}}, \ and\
  \bibinfo {author} {\bibfnamefont {L.~F.}\ \bibnamefont {Cugliandolo}},\
  }\href@noop {} {\bibfield  {journal} {\bibinfo  {journal} {Soft Matter}\
  }\textbf {\bibinfo {volume} {7}},\ \bibinfo {pages} {3726} (\bibinfo {year}
  {2011})}\BibitemShut {NoStop}%
\bibitem [{\citenamefont {Morozov}\ and\ \citenamefont
  {Pismen}(2010)}]{morozov2010motor}%
  \BibitemOpen
  \bibfield  {author} {\bibinfo {author} {\bibfnamefont {K.~I.}\ \bibnamefont
  {Morozov}}\ and\ \bibinfo {author} {\bibfnamefont {L.~M.}\ \bibnamefont
  {Pismen}},\ }\href@noop {} {\bibfield  {journal} {\bibinfo  {journal}
  {Physical Review E}\ }\textbf {\bibinfo {volume} {81}},\ \bibinfo {pages}
  {061922} (\bibinfo {year} {2010})}\BibitemShut {NoStop}%
\bibitem [{\citenamefont {Szamel}(2014)}]{szamel2014self}%
  \BibitemOpen
  \bibfield  {author} {\bibinfo {author} {\bibfnamefont {G.}~\bibnamefont
  {Szamel}},\ }\href@noop {} {\bibfield  {journal} {\bibinfo  {journal}
  {Physical Review E}\ }\textbf {\bibinfo {volume} {90}},\ \bibinfo {pages}
  {012111} (\bibinfo {year} {2014})}\BibitemShut {NoStop}%
\bibitem [{\citenamefont {Solon}\ \emph {et~al.}(2015)\citenamefont {Solon},
  \citenamefont {Cates},\ and\ \citenamefont {Tailleur}}]{solon2015active}%
  \BibitemOpen
  \bibfield  {author} {\bibinfo {author} {\bibfnamefont {A.~P.}\ \bibnamefont
  {Solon}}, \bibinfo {author} {\bibfnamefont {M.}~\bibnamefont {Cates}}, \ and\
  \bibinfo {author} {\bibfnamefont {J.}~\bibnamefont {Tailleur}},\ }\href@noop
  {} {\bibfield  {journal} {\bibinfo  {journal} {The European Physical Journal
  Special Topics}\ }\textbf {\bibinfo {volume} {224}},\ \bibinfo {pages} {1231}
  (\bibinfo {year} {2015})}\BibitemShut {NoStop}%
\bibitem [{\citenamefont {Chavanis}\ \emph {et~al.}(2004)\citenamefont
  {Chavanis}, \citenamefont {Ribot}, \citenamefont {Rosier},\ and\
  \citenamefont {Sire}}]{chavanis2004}%
  \BibitemOpen
  \bibfield  {author} {\bibinfo {author} {\bibfnamefont {P.}~\bibnamefont
  {Chavanis}}, \bibinfo {author} {\bibfnamefont {M.}~\bibnamefont {Ribot}},
  \bibinfo {author} {\bibfnamefont {C.}~\bibnamefont {Rosier}}, \ and\ \bibinfo
  {author} {\bibfnamefont {C.}~\bibnamefont {Sire}},\ }\href@noop {} {\bibfield
   {journal} {\bibinfo  {journal} {Banach Center Publ.}\ }\textbf {\bibinfo
  {volume} {66}},\ \bibinfo {pages} {287} (\bibinfo {year} {2004})}\BibitemShut
  {NoStop}%
\bibitem [{\citenamefont {Golestanian}(2012)}]{golestanian2012collective}%
  \BibitemOpen
  \bibfield  {author} {\bibinfo {author} {\bibfnamefont {R.}~\bibnamefont
  {Golestanian}},\ }\href@noop {} {\bibfield  {journal} {\bibinfo  {journal}
  {Physical Review Letters}\ }\textbf {\bibinfo {volume} {108}},\ \bibinfo
  {pages} {038303} (\bibinfo {year} {2012})}\BibitemShut {NoStop}%
\bibitem [{\citenamefont {Takatori}\ and\ \citenamefont
  {Brady}(2015)}]{takatori2015towards}%
  \BibitemOpen
  \bibfield  {author} {\bibinfo {author} {\bibfnamefont {S.~C.}\ \bibnamefont
  {Takatori}}\ and\ \bibinfo {author} {\bibfnamefont {J.~F.}\ \bibnamefont
  {Brady}},\ }\href@noop {} {\bibfield  {journal} {\bibinfo  {journal}
  {Physical Review E}\ }\textbf {\bibinfo {volume} {91}},\ \bibinfo {pages}
  {032117} (\bibinfo {year} {2015})}\BibitemShut {NoStop}%
\bibitem [{\citenamefont {Paliwal}\ \emph {et~al.}(2018)\citenamefont
  {Paliwal}, \citenamefont {Rodenburg}, \citenamefont {van Roij},\ and\
  \citenamefont {Dijkstra}}]{paliwal2018chemical}%
  \BibitemOpen
  \bibfield  {author} {\bibinfo {author} {\bibfnamefont {S.}~\bibnamefont
  {Paliwal}}, \bibinfo {author} {\bibfnamefont {J.}~\bibnamefont {Rodenburg}},
  \bibinfo {author} {\bibfnamefont {R.}~\bibnamefont {van Roij}}, \ and\
  \bibinfo {author} {\bibfnamefont {M.}~\bibnamefont {Dijkstra}},\ }\href@noop
  {} {\bibfield  {journal} {\bibinfo  {journal} {New Journal of Physics}\
  }\textbf {\bibinfo {volume} {20}},\ \bibinfo {pages} {015003} (\bibinfo
  {year} {2018})}\BibitemShut {NoStop}%
\bibitem [{\citenamefont {Thompson}\ \emph {et~al.}(2011)\citenamefont
  {Thompson}, \citenamefont {Tailleur}, \citenamefont {Cates},\ and\
  \citenamefont {Blythe}}]{thompson2011lattice}%
  \BibitemOpen
  \bibfield  {author} {\bibinfo {author} {\bibfnamefont {A.~G.}\ \bibnamefont
  {Thompson}}, \bibinfo {author} {\bibfnamefont {J.}~\bibnamefont {Tailleur}},
  \bibinfo {author} {\bibfnamefont {M.~E.}\ \bibnamefont {Cates}}, \ and\
  \bibinfo {author} {\bibfnamefont {R.~A.}\ \bibnamefont {Blythe}},\
  }\href@noop {} {\bibfield  {journal} {\bibinfo  {journal} {Journal of
  Statistical Mechanics: Theory and Experiment}\ }\textbf {\bibinfo {volume}
  {2011}},\ \bibinfo {pages} {P02029} (\bibinfo {year} {2011})}\BibitemShut
  {NoStop}%
\bibitem [{\citenamefont {Bi}\ \emph {et~al.}(2016)\citenamefont {Bi},
  \citenamefont {Yang}, \citenamefont {Marchetti},\ and\ \citenamefont
  {Manning}}]{bi2016motility}%
  \BibitemOpen
  \bibfield  {author} {\bibinfo {author} {\bibfnamefont {D.}~\bibnamefont
  {Bi}}, \bibinfo {author} {\bibfnamefont {X.}~\bibnamefont {Yang}}, \bibinfo
  {author} {\bibfnamefont {M.~C.}\ \bibnamefont {Marchetti}}, \ and\ \bibinfo
  {author} {\bibfnamefont {M.~L.}\ \bibnamefont {Manning}},\ }\href@noop {}
  {\bibfield  {journal} {\bibinfo  {journal} {Physical Review X}\ }\textbf
  {\bibinfo {volume} {6}},\ \bibinfo {pages} {021011} (\bibinfo {year}
  {2016})}\BibitemShut {NoStop}%
\bibitem [{\citenamefont {Grafke}\ \emph {et~al.}(2017)\citenamefont {Grafke},
  \citenamefont {Cates},\ and\ \citenamefont
  {Vanden-Eijnden}}]{grafke2017spatiotemporal}%
  \BibitemOpen
  \bibfield  {author} {\bibinfo {author} {\bibfnamefont {T.}~\bibnamefont
  {Grafke}}, \bibinfo {author} {\bibfnamefont {M.~E.}\ \bibnamefont {Cates}}, \
  and\ \bibinfo {author} {\bibfnamefont {E.}~\bibnamefont {Vanden-Eijnden}},\
  }\href@noop {} {\bibfield  {journal} {\bibinfo  {journal} {Physical Review
  Letters}\ }\textbf {\bibinfo {volume} {119}},\ \bibinfo {pages} {188003}
  (\bibinfo {year} {2017})}\BibitemShut {NoStop}%
\bibitem [{\citenamefont {Nardini}\ \emph {et~al.}(2017)\citenamefont
  {Nardini}, \citenamefont {Fodor}, \citenamefont {Tjhung}, \citenamefont
  {Van~Wijland}, \citenamefont {Tailleur},\ and\ \citenamefont
  {Cates}}]{nardini2017entropy}%
  \BibitemOpen
  \bibfield  {author} {\bibinfo {author} {\bibfnamefont {C.}~\bibnamefont
  {Nardini}}, \bibinfo {author} {\bibfnamefont {{\'E}.}~\bibnamefont {Fodor}},
  \bibinfo {author} {\bibfnamefont {E.}~\bibnamefont {Tjhung}}, \bibinfo
  {author} {\bibfnamefont {F.}~\bibnamefont {Van~Wijland}}, \bibinfo {author}
  {\bibfnamefont {J.}~\bibnamefont {Tailleur}}, \ and\ \bibinfo {author}
  {\bibfnamefont {M.~E.}\ \bibnamefont {Cates}},\ }\href@noop {} {\bibfield
  {journal} {\bibinfo  {journal} {Physical Review X}\ }\textbf {\bibinfo
  {volume} {7}},\ \bibinfo {pages} {021007} (\bibinfo {year}
  {2017})}\BibitemShut {NoStop}%
\bibitem [{Note1()}]{Note1}%
  \BibitemOpen
  \bibinfo {note} {Taxis describes the biased motion of an entity in response
  to an external signal}\BibitemShut {NoStop}%
\bibitem [{\citenamefont {Berg}(2008)}]{berg2008coli}%
  \BibitemOpen
  \bibfield  {author} {\bibinfo {author} {\bibfnamefont {H.~C.}\ \bibnamefont
  {Berg}},\ }\href@noop {} {\emph {\bibinfo {title} {E. coli in Motion}}}\
  (\bibinfo  {publisher} {Springer Science \& Business Media},\ \bibinfo {year}
  {2008})\BibitemShut {NoStop}%
\bibitem [{\citenamefont {Polin}\ \emph {et~al.}(2009)\citenamefont {Polin},
  \citenamefont {Tuval}, \citenamefont {Drescher}, \citenamefont {Gollub},\
  and\ \citenamefont {Goldstein}}]{polin2009chlamydomonas}%
  \BibitemOpen
  \bibfield  {author} {\bibinfo {author} {\bibfnamefont {M.}~\bibnamefont
  {Polin}}, \bibinfo {author} {\bibfnamefont {I.}~\bibnamefont {Tuval}},
  \bibinfo {author} {\bibfnamefont {K.}~\bibnamefont {Drescher}}, \bibinfo
  {author} {\bibfnamefont {J.~P.}\ \bibnamefont {Gollub}}, \ and\ \bibinfo
  {author} {\bibfnamefont {R.~E.}\ \bibnamefont {Goldstein}},\ }\href@noop {}
  {\bibfield  {journal} {\bibinfo  {journal} {Science}\ }\textbf {\bibinfo
  {volume} {325}},\ \bibinfo {pages} {487} (\bibinfo {year}
  {2009})}\BibitemShut {NoStop}%
\bibitem [{\citenamefont {Sunyer}\ \emph {et~al.}(2016)\citenamefont {Sunyer},
  \citenamefont {Conte}, \citenamefont {Escribano}, \citenamefont
  {Elosegui-Artola}, \citenamefont {Labernadie}, \citenamefont {Valon},
  \citenamefont {Navajas}, \citenamefont {Garc{\'\i}a-Aznar}, \citenamefont
  {Mu{\~n}oz}, \citenamefont {Roca-Cusachs} \emph
  {et~al.}}]{sunyer2016collective}%
  \BibitemOpen
  \bibfield  {author} {\bibinfo {author} {\bibfnamefont {R.}~\bibnamefont
  {Sunyer}}, \bibinfo {author} {\bibfnamefont {V.}~\bibnamefont {Conte}},
  \bibinfo {author} {\bibfnamefont {J.}~\bibnamefont {Escribano}}, \bibinfo
  {author} {\bibfnamefont {A.}~\bibnamefont {Elosegui-Artola}}, \bibinfo
  {author} {\bibfnamefont {A.}~\bibnamefont {Labernadie}}, \bibinfo {author}
  {\bibfnamefont {L.}~\bibnamefont {Valon}}, \bibinfo {author} {\bibfnamefont
  {D.}~\bibnamefont {Navajas}}, \bibinfo {author} {\bibfnamefont {J.~M.}\
  \bibnamefont {Garc{\'\i}a-Aznar}}, \bibinfo {author} {\bibfnamefont {J.~J.}\
  \bibnamefont {Mu{\~n}oz}}, \bibinfo {author} {\bibfnamefont {P.}~\bibnamefont
  {Roca-Cusachs}},  \emph {et~al.},\ }\href@noop {} {\bibfield  {journal}
  {\bibinfo  {journal} {Science}\ }\textbf {\bibinfo {volume} {353}},\ \bibinfo
  {pages} {1157} (\bibinfo {year} {2016})}\BibitemShut {NoStop}%
\bibitem [{\citenamefont {Soto}\ and\ \citenamefont
  {Golestanian}(2014)}]{soto2014self}%
  \BibitemOpen
  \bibfield  {author} {\bibinfo {author} {\bibfnamefont {R.}~\bibnamefont
  {Soto}}\ and\ \bibinfo {author} {\bibfnamefont {R.}~\bibnamefont
  {Golestanian}},\ }\href@noop {} {\bibfield  {journal} {\bibinfo  {journal}
  {Physical Review Letters}\ }\textbf {\bibinfo {volume} {112}},\ \bibinfo
  {pages} {068301} (\bibinfo {year} {2014})}\BibitemShut {NoStop}%
\bibitem [{\citenamefont {Pohl}\ and\ \citenamefont
  {Stark}(2014)}]{pohl2014dynamic}%
  \BibitemOpen
  \bibfield  {author} {\bibinfo {author} {\bibfnamefont {O.}~\bibnamefont
  {Pohl}}\ and\ \bibinfo {author} {\bibfnamefont {H.}~\bibnamefont {Stark}},\
  }\href@noop {} {\bibfield  {journal} {\bibinfo  {journal} {Physical Review
  Letters}\ }\textbf {\bibinfo {volume} {112}},\ \bibinfo {pages} {238303}
  (\bibinfo {year} {2014})}\BibitemShut {NoStop}%
\bibitem [{\citenamefont {Woodward}\ \emph {et~al.}(1995)\citenamefont
  {Woodward}, \citenamefont {Tyson}, \citenamefont {Myerscough}, \citenamefont
  {Murray}, \citenamefont {Budrene},\ and\ \citenamefont
  {Berg}}]{woodward1995spatio}%
  \BibitemOpen
  \bibfield  {author} {\bibinfo {author} {\bibfnamefont {D.}~\bibnamefont
  {Woodward}}, \bibinfo {author} {\bibfnamefont {R.}~\bibnamefont {Tyson}},
  \bibinfo {author} {\bibfnamefont {M.}~\bibnamefont {Myerscough}}, \bibinfo
  {author} {\bibfnamefont {J.}~\bibnamefont {Murray}}, \bibinfo {author}
  {\bibfnamefont {E.}~\bibnamefont {Budrene}}, \ and\ \bibinfo {author}
  {\bibfnamefont {H.}~\bibnamefont {Berg}},\ }\href@noop {} {\bibfield
  {journal} {\bibinfo  {journal} {Biophysical journal}\ }\textbf {\bibinfo
  {volume} {68}},\ \bibinfo {pages} {2181} (\bibinfo {year}
  {1995})}\BibitemShut {NoStop}%
\bibitem [{\citenamefont {Brenner}\ \emph {et~al.}(1998)\citenamefont
  {Brenner}, \citenamefont {Levitov},\ and\ \citenamefont
  {Budrene}}]{brenner1998physical}%
  \BibitemOpen
  \bibfield  {author} {\bibinfo {author} {\bibfnamefont {M.~P.}\ \bibnamefont
  {Brenner}}, \bibinfo {author} {\bibfnamefont {L.~S.}\ \bibnamefont
  {Levitov}}, \ and\ \bibinfo {author} {\bibfnamefont {E.~O.}\ \bibnamefont
  {Budrene}},\ }\href@noop {} {\bibfield  {journal} {\bibinfo  {journal}
  {Biophysical journal}\ }\textbf {\bibinfo {volume} {74}},\ \bibinfo {pages}
  {1677} (\bibinfo {year} {1998})}\BibitemShut {NoStop}%
\bibitem [{\citenamefont {Chatterjee}\ \emph {et~al.}(2011)\citenamefont
  {Chatterjee}, \citenamefont {da~Silveira},\ and\ \citenamefont
  {Kafri}}]{chatterjee2011chemotaxis}%
  \BibitemOpen
  \bibfield  {author} {\bibinfo {author} {\bibfnamefont {S.}~\bibnamefont
  {Chatterjee}}, \bibinfo {author} {\bibfnamefont {R.~A.}\ \bibnamefont
  {da~Silveira}}, \ and\ \bibinfo {author} {\bibfnamefont {Y.}~\bibnamefont
  {Kafri}},\ }\href@noop {} {\bibfield  {journal} {\bibinfo  {journal} {PLoS
  computational biology}\ }\textbf {\bibinfo {volume} {7}} (\bibinfo {year}
  {2011})}\BibitemShut {NoStop}%
\bibitem [{\citenamefont {Saragosti}\ \emph {et~al.}(2011)\citenamefont
  {Saragosti}, \citenamefont {Calvez}, \citenamefont {Bournaveas},
  \citenamefont {Perthame}, \citenamefont {Buguin},\ and\ \citenamefont
  {Silberzan}}]{saragosti2011directional}%
  \BibitemOpen
  \bibfield  {author} {\bibinfo {author} {\bibfnamefont {J.}~\bibnamefont
  {Saragosti}}, \bibinfo {author} {\bibfnamefont {V.}~\bibnamefont {Calvez}},
  \bibinfo {author} {\bibfnamefont {N.}~\bibnamefont {Bournaveas}}, \bibinfo
  {author} {\bibfnamefont {B.}~\bibnamefont {Perthame}}, \bibinfo {author}
  {\bibfnamefont {A.}~\bibnamefont {Buguin}}, \ and\ \bibinfo {author}
  {\bibfnamefont {P.}~\bibnamefont {Silberzan}},\ }\href@noop {} {\bibfield
  {journal} {\bibinfo  {journal} {Proceedings of the National Academy of
  Sciences}\ }\textbf {\bibinfo {volume} {108}},\ \bibinfo {pages} {16235}
  (\bibinfo {year} {2011})}\BibitemShut {NoStop}%
\bibitem [{\citenamefont {Saha}\ \emph {et~al.}(2014)\citenamefont {Saha},
  \citenamefont {Golestanian},\ and\ \citenamefont
  {Ramaswamy}}]{saha2014clusters}%
  \BibitemOpen
  \bibfield  {author} {\bibinfo {author} {\bibfnamefont {S.}~\bibnamefont
  {Saha}}, \bibinfo {author} {\bibfnamefont {R.}~\bibnamefont {Golestanian}}, \
  and\ \bibinfo {author} {\bibfnamefont {S.}~\bibnamefont {Ramaswamy}},\
  }\href@noop {} {\bibfield  {journal} {\bibinfo  {journal} {Physical Review
  E}\ }\textbf {\bibinfo {volume} {89}},\ \bibinfo {pages} {062316} (\bibinfo
  {year} {2014})}\BibitemShut {NoStop}%
\bibitem [{\citenamefont {Agudo-Canalejo}\ and\ \citenamefont
  {Golestanian}(2019)}]{agudo2019active}%
  \BibitemOpen
  \bibfield  {author} {\bibinfo {author} {\bibfnamefont {J.}~\bibnamefont
  {Agudo-Canalejo}}\ and\ \bibinfo {author} {\bibfnamefont {R.}~\bibnamefont
  {Golestanian}},\ }\href@noop {} {\bibfield  {journal} {\bibinfo  {journal}
  {Physical Review Letters}\ }\textbf {\bibinfo {volume} {123}},\ \bibinfo
  {pages} {018101} (\bibinfo {year} {2019})}\BibitemShut {NoStop}%
\bibitem [{\citenamefont {Nasouri}\ and\ \citenamefont
  {Golestanian}(2020)}]{nasouri2020exact}%
  \BibitemOpen
  \bibfield  {author} {\bibinfo {author} {\bibfnamefont {B.}~\bibnamefont
  {Nasouri}}\ and\ \bibinfo {author} {\bibfnamefont {R.}~\bibnamefont
  {Golestanian}},\ }\href@noop {} {\bibfield  {journal} {\bibinfo  {journal}
  {Physical Review Letters}\ }\textbf {\bibinfo {volume} {124}},\ \bibinfo
  {pages} {168003} (\bibinfo {year} {2020})}\BibitemShut {NoStop}%
\bibitem [{\citenamefont {Goldstein}(1996)}]{goldstein1996traveling}%
  \BibitemOpen
  \bibfield  {author} {\bibinfo {author} {\bibfnamefont {R.~E.}\ \bibnamefont
  {Goldstein}},\ }\href@noop {} {\bibfield  {journal} {\bibinfo  {journal}
  {Physical Review Letters}\ }\textbf {\bibinfo {volume} {77}},\ \bibinfo
  {pages} {775} (\bibinfo {year} {1996})}\BibitemShut {NoStop}%
\bibitem [{\citenamefont {Seyrich}\ \emph {et~al.}(2019)\citenamefont
  {Seyrich}, \citenamefont {Palugniok},\ and\ \citenamefont
  {Stark}}]{seyrich2019traveling}%
  \BibitemOpen
  \bibfield  {author} {\bibinfo {author} {\bibfnamefont {M.}~\bibnamefont
  {Seyrich}}, \bibinfo {author} {\bibfnamefont {A.}~\bibnamefont {Palugniok}},
  \ and\ \bibinfo {author} {\bibfnamefont {H.}~\bibnamefont {Stark}},\
  }\href@noop {} {\bibfield  {journal} {\bibinfo  {journal} {New Journal of
  Physics}\ }\textbf {\bibinfo {volume} {21}},\ \bibinfo {pages} {103001}
  (\bibinfo {year} {2019})}\BibitemShut {NoStop}%
\bibitem [{\citenamefont {Mahdisoltani}\ \emph {et~al.}(2019)\citenamefont
  {Mahdisoltani}, \citenamefont {Zinati}, \citenamefont {Duclut}, \citenamefont
  {Gambassi},\ and\ \citenamefont {Golestanian}}]{mahdisoltani2019controlled}%
  \BibitemOpen
  \bibfield  {author} {\bibinfo {author} {\bibfnamefont {S.}~\bibnamefont
  {Mahdisoltani}}, \bibinfo {author} {\bibfnamefont {R.~B.~A.}\ \bibnamefont
  {Zinati}}, \bibinfo {author} {\bibfnamefont {C.}~\bibnamefont {Duclut}},
  \bibinfo {author} {\bibfnamefont {A.}~\bibnamefont {Gambassi}}, \ and\
  \bibinfo {author} {\bibfnamefont {R.}~\bibnamefont {Golestanian}},\
  }\href@noop {} {\bibfield  {journal} {\bibinfo  {journal} {arXiv preprint
  arXiv:1911.08115}\ } (\bibinfo {year} {2019})}\BibitemShut {NoStop}%
\bibitem [{Note2()}]{Note2}%
  \BibitemOpen
  \bibinfo {note} {The rotational diffusion is understood as an It\=o-Langevin
  process.}\BibitemShut {Stop}%
\bibitem [{\citenamefont {B{\"a}uerle}\ \emph {et~al.}(2018)\citenamefont
  {B{\"a}uerle}, \citenamefont {Fischer}, \citenamefont {Speck},\ and\
  \citenamefont {Bechinger}}]{bauerle2018self}%
  \BibitemOpen
  \bibfield  {author} {\bibinfo {author} {\bibfnamefont {T.}~\bibnamefont
  {B{\"a}uerle}}, \bibinfo {author} {\bibfnamefont {A.}~\bibnamefont
  {Fischer}}, \bibinfo {author} {\bibfnamefont {T.}~\bibnamefont {Speck}}, \
  and\ \bibinfo {author} {\bibfnamefont {C.}~\bibnamefont {Bechinger}},\
  }\href@noop {} {\bibfield  {journal} {\bibinfo  {journal} {Nature
  communications}\ }\textbf {\bibinfo {volume} {9}},\ \bibinfo {pages} {1}
  (\bibinfo {year} {2018})}\BibitemShut {NoStop}%
\bibitem [{\citenamefont {Lavergne}\ \emph {et~al.}(2019)\citenamefont
  {Lavergne}, \citenamefont {Wendehenne}, \citenamefont {B{\"a}uerle},\ and\
  \citenamefont {Bechinger}}]{lavergne2019group}%
  \BibitemOpen
  \bibfield  {author} {\bibinfo {author} {\bibfnamefont {F.~A.}\ \bibnamefont
  {Lavergne}}, \bibinfo {author} {\bibfnamefont {H.}~\bibnamefont
  {Wendehenne}}, \bibinfo {author} {\bibfnamefont {T.}~\bibnamefont
  {B{\"a}uerle}}, \ and\ \bibinfo {author} {\bibfnamefont {C.}~\bibnamefont
  {Bechinger}},\ }\href@noop {} {\bibfield  {journal} {\bibinfo  {journal}
  {Science}\ }\textbf {\bibinfo {volume} {364}},\ \bibinfo {pages} {70}
  (\bibinfo {year} {2019})}\BibitemShut {NoStop}%
\bibitem [{\citenamefont {Buttinoni}\ \emph {et~al.}(2013)\citenamefont
  {Buttinoni}, \citenamefont {Bialk{\'e}}, \citenamefont {K{\"u}mmel},
  \citenamefont {L{\"o}wen}, \citenamefont {Bechinger},\ and\ \citenamefont
  {Speck}}]{buttinoni2013dynamical}%
  \BibitemOpen
  \bibfield  {author} {\bibinfo {author} {\bibfnamefont {I.}~\bibnamefont
  {Buttinoni}}, \bibinfo {author} {\bibfnamefont {J.}~\bibnamefont
  {Bialk{\'e}}}, \bibinfo {author} {\bibfnamefont {F.}~\bibnamefont
  {K{\"u}mmel}}, \bibinfo {author} {\bibfnamefont {H.}~\bibnamefont
  {L{\"o}wen}}, \bibinfo {author} {\bibfnamefont {C.}~\bibnamefont
  {Bechinger}}, \ and\ \bibinfo {author} {\bibfnamefont {T.}~\bibnamefont
  {Speck}},\ }\href@noop {} {\bibfield  {journal} {\bibinfo  {journal}
  {Physical Review Letters}\ }\textbf {\bibinfo {volume} {110}},\ \bibinfo
  {pages} {238301} (\bibinfo {year} {2013})}\BibitemShut {NoStop}%
\bibitem [{\citenamefont {Vizsnyiczai}\ \emph {et~al.}(2017)\citenamefont
  {Vizsnyiczai}, \citenamefont {Frangipane}, \citenamefont {Maggi},
  \citenamefont {Saglimbeni}, \citenamefont {Bianchi},\ and\ \citenamefont
  {Di~Leonardo}}]{vizsnyiczai2017light}%
  \BibitemOpen
  \bibfield  {author} {\bibinfo {author} {\bibfnamefont {G.}~\bibnamefont
  {Vizsnyiczai}}, \bibinfo {author} {\bibfnamefont {G.}~\bibnamefont
  {Frangipane}}, \bibinfo {author} {\bibfnamefont {C.}~\bibnamefont {Maggi}},
  \bibinfo {author} {\bibfnamefont {F.}~\bibnamefont {Saglimbeni}}, \bibinfo
  {author} {\bibfnamefont {S.}~\bibnamefont {Bianchi}}, \ and\ \bibinfo
  {author} {\bibfnamefont {R.}~\bibnamefont {Di~Leonardo}},\ }\href@noop {}
  {\bibfield  {journal} {\bibinfo  {journal} {Nature communications}\ }\textbf
  {\bibinfo {volume} {8}},\ \bibinfo {pages} {1} (\bibinfo {year}
  {2017})}\BibitemShut {NoStop}%
\bibitem [{\citenamefont {Frangipane}\ \emph {et~al.}(2018)\citenamefont
  {Frangipane}, \citenamefont {Dell'Arciprete}, \citenamefont {Petracchini},
  \citenamefont {Maggi}, \citenamefont {Saglimbeni}, \citenamefont {Bianchi},
  \citenamefont {Vizsnyiczai}, \citenamefont {Bernardini},\ and\ \citenamefont
  {Di~Leonardo}}]{frangipane2018dynamic}%
  \BibitemOpen
  \bibfield  {author} {\bibinfo {author} {\bibfnamefont {G.}~\bibnamefont
  {Frangipane}}, \bibinfo {author} {\bibfnamefont {D.}~\bibnamefont
  {Dell'Arciprete}}, \bibinfo {author} {\bibfnamefont {S.}~\bibnamefont
  {Petracchini}}, \bibinfo {author} {\bibfnamefont {C.}~\bibnamefont {Maggi}},
  \bibinfo {author} {\bibfnamefont {F.}~\bibnamefont {Saglimbeni}}, \bibinfo
  {author} {\bibfnamefont {S.}~\bibnamefont {Bianchi}}, \bibinfo {author}
  {\bibfnamefont {G.}~\bibnamefont {Vizsnyiczai}}, \bibinfo {author}
  {\bibfnamefont {M.~L.}\ \bibnamefont {Bernardini}}, \ and\ \bibinfo {author}
  {\bibfnamefont {R.}~\bibnamefont {Di~Leonardo}},\ }\href@noop {} {\bibfield
  {journal} {\bibinfo  {journal} {Elife}\ }\textbf {\bibinfo {volume} {7}},\
  \bibinfo {pages} {e36608} (\bibinfo {year} {2018})}\BibitemShut {NoStop}%
\bibitem [{\citenamefont {Arlt}\ \emph {et~al.}(2018)\citenamefont {Arlt},
  \citenamefont {Martinez}, \citenamefont {Dawson}, \citenamefont {Pilizota},\
  and\ \citenamefont {Poon}}]{arlt2018painting}%
  \BibitemOpen
  \bibfield  {author} {\bibinfo {author} {\bibfnamefont {J.}~\bibnamefont
  {Arlt}}, \bibinfo {author} {\bibfnamefont {V.~A.}\ \bibnamefont {Martinez}},
  \bibinfo {author} {\bibfnamefont {A.}~\bibnamefont {Dawson}}, \bibinfo
  {author} {\bibfnamefont {T.}~\bibnamefont {Pilizota}}, \ and\ \bibinfo
  {author} {\bibfnamefont {W.~C.}\ \bibnamefont {Poon}},\ }\href@noop {}
  {\bibfield  {journal} {\bibinfo  {journal} {Nature communications}\ }\textbf
  {\bibinfo {volume} {9}},\ \bibinfo {pages} {1} (\bibinfo {year}
  {2018})}\BibitemShut {NoStop}%
\bibitem [{\citenamefont {Arlt}\ \emph {et~al.}(2019)\citenamefont {Arlt},
  \citenamefont {Martinez}, \citenamefont {Dawson}, \citenamefont {Pilizota},\
  and\ \citenamefont {Poon}}]{arlt2019dynamics}%
  \BibitemOpen
  \bibfield  {author} {\bibinfo {author} {\bibfnamefont {J.}~\bibnamefont
  {Arlt}}, \bibinfo {author} {\bibfnamefont {V.~A.}\ \bibnamefont {Martinez}},
  \bibinfo {author} {\bibfnamefont {A.}~\bibnamefont {Dawson}}, \bibinfo
  {author} {\bibfnamefont {T.}~\bibnamefont {Pilizota}}, \ and\ \bibinfo
  {author} {\bibfnamefont {W.~C.}\ \bibnamefont {Poon}},\ }\href@noop {}
  {\bibfield  {journal} {\bibinfo  {journal} {Nature communications}\ }\textbf
  {\bibinfo {volume} {10}},\ \bibinfo {pages} {1} (\bibinfo {year}
  {2019})}\BibitemShut {NoStop}%
\bibitem [{\citenamefont {Schnitzer}(1993)}]{schnitzer1993theory}%
  \BibitemOpen
  \bibfield  {author} {\bibinfo {author} {\bibfnamefont {M.~J.}\ \bibnamefont
  {Schnitzer}},\ }\href@noop {} {\bibfield  {journal} {\bibinfo  {journal}
  {Physical Review E}\ }\textbf {\bibinfo {volume} {48}},\ \bibinfo {pages}
  {2553} (\bibinfo {year} {1993})}\BibitemShut {NoStop}%
\bibitem [{\citenamefont {Bertin}\ \emph {et~al.}(2006)\citenamefont {Bertin},
  \citenamefont {Droz},\ and\ \citenamefont
  {Gr{\'e}goire}}]{bertin2006boltzmann}%
  \BibitemOpen
  \bibfield  {author} {\bibinfo {author} {\bibfnamefont {E.}~\bibnamefont
  {Bertin}}, \bibinfo {author} {\bibfnamefont {M.}~\bibnamefont {Droz}}, \ and\
  \bibinfo {author} {\bibfnamefont {G.}~\bibnamefont {Gr{\'e}goire}},\
  }\href@noop {} {\bibfield  {journal} {\bibinfo  {journal} {Physical Review
  E}\ }\textbf {\bibinfo {volume} {74}},\ \bibinfo {pages} {022101} (\bibinfo
  {year} {2006})}\BibitemShut {NoStop}%
\bibitem [{\citenamefont {Liebchen}\ \emph {et~al.}(2017)\citenamefont
  {Liebchen}, \citenamefont {Marenduzzo},\ and\ \citenamefont
  {Cates}}]{liebchen2017phoretic}%
  \BibitemOpen
  \bibfield  {author} {\bibinfo {author} {\bibfnamefont {B.}~\bibnamefont
  {Liebchen}}, \bibinfo {author} {\bibfnamefont {D.}~\bibnamefont
  {Marenduzzo}}, \ and\ \bibinfo {author} {\bibfnamefont {M.~E.}\ \bibnamefont
  {Cates}},\ }\href@noop {} {\bibfield  {journal} {\bibinfo  {journal}
  {Physical Review Letters}\ }\textbf {\bibinfo {volume} {118}},\ \bibinfo
  {pages} {268001} (\bibinfo {year} {2017})}\BibitemShut {NoStop}%
\bibitem [{\citenamefont {Kourbane-Houssene}\ \emph {et~al.}(2018)\citenamefont
  {Kourbane-Houssene}, \citenamefont {Erignoux}, \citenamefont {Bodineau},\
  and\ \citenamefont {Tailleur}}]{kourbane2018exact}%
  \BibitemOpen
  \bibfield  {author} {\bibinfo {author} {\bibfnamefont {M.}~\bibnamefont
  {Kourbane-Houssene}}, \bibinfo {author} {\bibfnamefont {C.}~\bibnamefont
  {Erignoux}}, \bibinfo {author} {\bibfnamefont {T.}~\bibnamefont {Bodineau}},
  \ and\ \bibinfo {author} {\bibfnamefont {J.}~\bibnamefont {Tailleur}},\
  }\href@noop {} {\bibfield  {journal} {\bibinfo  {journal} {Physical Review
  Letters}\ }\textbf {\bibinfo {volume} {120}},\ \bibinfo {pages} {268003}
  (\bibinfo {year} {2018})}\BibitemShut {NoStop}%
\bibitem [{sup()}]{supp}%
  \BibitemOpen
  \href@noop {} {}\bibinfo {note} {See Supplemental Material [url], which
  includes theoretical and numerical details, as well as Refs. XXX}\BibitemShut
  {NoStop}%
\bibitem [{\citenamefont {Chaikin}\ \emph {et~al.}(1995)\citenamefont
  {Chaikin}, \citenamefont {Lubensky},\ and\ \citenamefont
  {Witten}}]{chaikin1995principles}%
  \BibitemOpen
  \bibfield  {author} {\bibinfo {author} {\bibfnamefont {P.~M.}\ \bibnamefont
  {Chaikin}}, \bibinfo {author} {\bibfnamefont {T.~C.}\ \bibnamefont
  {Lubensky}}, \ and\ \bibinfo {author} {\bibfnamefont {T.~A.}\ \bibnamefont
  {Witten}},\ }\href@noop {} {\emph {\bibinfo {title} {Principles of condensed
  matter physics}}},\ Vol.~\bibinfo {volume} {10}\ (\bibinfo  {publisher}
  {Cambridge university press Cambridge},\ \bibinfo {year} {1995})\BibitemShut
  {NoStop}%
\bibitem [{\citenamefont {Anderson}(1989)}]{anderson1989variational}%
  \BibitemOpen
  \bibfield  {author} {\bibinfo {author} {\bibfnamefont {I.~M.}\ \bibnamefont
  {Anderson}},\ }\href@noop {} {\emph {\bibinfo {title} {The variational
  bicomplex}}},\ \bibinfo {type} {Tech. Rep.}\ (\bibinfo  {institution} {Utah
  State Technical Report, 1989, http://math. usu. edu/~ fg mp},\ \bibinfo
  {year} {1989})\BibitemShut {NoStop}%
\bibitem [{Note3()}]{Note3}%
  \BibitemOpen
  \bibinfo {note} {As usual, there are topological requirement for the
  conditions to be sufficient~\cite {anderson1989variational}}\BibitemShut
  {NoStop}%
\bibitem [{\citenamefont {Chavanis}(2010)}]{chavanis2010stochastic}%
  \BibitemOpen
  \bibfield  {author} {\bibinfo {author} {\bibfnamefont {P.-H.}\ \bibnamefont
  {Chavanis}},\ }\href@noop {} {\bibfield  {journal} {\bibinfo  {journal}
  {Communications in Nonlinear Science and Numerical Simulation}\ }\textbf
  {\bibinfo {volume} {15}},\ \bibinfo {pages} {60} (\bibinfo {year}
  {2010})}\BibitemShut {NoStop}%
\bibitem [{\citenamefont {Malescio}\ and\ \citenamefont
  {Pellicane}(2003)}]{malescio2003stripe}%
  \BibitemOpen
  \bibfield  {author} {\bibinfo {author} {\bibfnamefont {G.}~\bibnamefont
  {Malescio}}\ and\ \bibinfo {author} {\bibfnamefont {G.}~\bibnamefont
  {Pellicane}},\ }\href@noop {} {\bibfield  {journal} {\bibinfo  {journal}
  {Nature materials}\ }\textbf {\bibinfo {volume} {2}},\ \bibinfo {pages} {97}
  (\bibinfo {year} {2003})}\BibitemShut {NoStop}%
\bibitem [{\citenamefont {Glaser}\ \emph {et~al.}(2007)\citenamefont {Glaser},
  \citenamefont {Grason}, \citenamefont {Kamien}, \citenamefont
  {Ko{\v{s}}mrlj}, \citenamefont {Santangelo},\ and\ \citenamefont
  {Ziherl}}]{glaser2007soft}%
  \BibitemOpen
  \bibfield  {author} {\bibinfo {author} {\bibfnamefont {M.~A.}\ \bibnamefont
  {Glaser}}, \bibinfo {author} {\bibfnamefont {G.~M.}\ \bibnamefont {Grason}},
  \bibinfo {author} {\bibfnamefont {R.~D.}\ \bibnamefont {Kamien}}, \bibinfo
  {author} {\bibfnamefont {A.}~\bibnamefont {Ko{\v{s}}mrlj}}, \bibinfo {author}
  {\bibfnamefont {C.~D.}\ \bibnamefont {Santangelo}}, \ and\ \bibinfo {author}
  {\bibfnamefont {P.}~\bibnamefont {Ziherl}},\ }\href@noop {} {\bibfield
  {journal} {\bibinfo  {journal} {EPL (Europhysics Letters)}\ }\textbf
  {\bibinfo {volume} {78}},\ \bibinfo {pages} {46004} (\bibinfo {year}
  {2007})}\BibitemShut {NoStop}%
\bibitem [{\citenamefont {L{\"o}wen}(2011)}]{lowen2011applications}%
  \BibitemOpen
  \bibfield  {author} {\bibinfo {author} {\bibfnamefont {H.}~\bibnamefont
  {L{\"o}wen}},\ }in\ \href@noop {} {\emph {\bibinfo {booktitle} {Understanding
  Soft Condensed Matter Via Modeling And Computation}}}\ (\bibinfo  {publisher}
  {World Scientific},\ \bibinfo {year} {2011})\ pp.\ \bibinfo {pages}
  {9--45}\BibitemShut {NoStop}%
\bibitem [{\citenamefont {Ruelle}(1999)}]{ruelle1999statistical}%
  \BibitemOpen
  \bibfield  {author} {\bibinfo {author} {\bibfnamefont {D.}~\bibnamefont
  {Ruelle}},\ }\href@noop {} {\emph {\bibinfo {title} {Statistical mechanics:
  Rigorous results}}}\ (\bibinfo  {publisher} {World Scientific},\ \bibinfo
  {year} {1999})\BibitemShut {NoStop}%
\bibitem [{\citenamefont {Othmer}\ and\ \citenamefont
  {Hillen}(2000)}]{othmer2000diffusion}%
  \BibitemOpen
  \bibfield  {author} {\bibinfo {author} {\bibfnamefont {H.~G.}\ \bibnamefont
  {Othmer}}\ and\ \bibinfo {author} {\bibfnamefont {T.}~\bibnamefont
  {Hillen}},\ }\href@noop {} {\bibfield  {journal} {\bibinfo  {journal} {SIAM
  Journal on Applied Mathematics}\ }\textbf {\bibinfo {volume} {61}},\ \bibinfo
  {pages} {751} (\bibinfo {year} {2000})}\BibitemShut {NoStop}%
\bibitem [{\citenamefont {Marconi}\ \emph
  {et~al.}(2016{\natexlab{b}})\citenamefont {Marconi}, \citenamefont {Gnan},
  \citenamefont {Paoluzzi}, \citenamefont {Maggi},\ and\ \citenamefont
  {Di~Leonardo}}]{marconi2016velocity}%
  \BibitemOpen
  \bibfield  {author} {\bibinfo {author} {\bibfnamefont {U.~M.~B.}\
  \bibnamefont {Marconi}}, \bibinfo {author} {\bibfnamefont {N.}~\bibnamefont
  {Gnan}}, \bibinfo {author} {\bibfnamefont {M.}~\bibnamefont {Paoluzzi}},
  \bibinfo {author} {\bibfnamefont {C.}~\bibnamefont {Maggi}}, \ and\ \bibinfo
  {author} {\bibfnamefont {R.}~\bibnamefont {Di~Leonardo}},\ }\href@noop {}
  {\bibfield  {journal} {\bibinfo  {journal} {Scientific reports}\ }\textbf
  {\bibinfo {volume} {6}},\ \bibinfo {pages} {1} (\bibinfo {year}
  {2016}{\natexlab{b}})}\BibitemShut {NoStop}%
\bibitem [{\citenamefont {Fodor}\ \emph {et~al.}(2016)\citenamefont {Fodor},
  \citenamefont {Nardini}, \citenamefont {Cates}, \citenamefont {Tailleur},
  \citenamefont {Visco},\ and\ \citenamefont {van Wijland}}]{fodor2016far}%
  \BibitemOpen
  \bibfield  {author} {\bibinfo {author} {\bibfnamefont {{\'E}.}~\bibnamefont
  {Fodor}}, \bibinfo {author} {\bibfnamefont {C.}~\bibnamefont {Nardini}},
  \bibinfo {author} {\bibfnamefont {M.~E.}\ \bibnamefont {Cates}}, \bibinfo
  {author} {\bibfnamefont {J.}~\bibnamefont {Tailleur}}, \bibinfo {author}
  {\bibfnamefont {P.}~\bibnamefont {Visco}}, \ and\ \bibinfo {author}
  {\bibfnamefont {F.}~\bibnamefont {van Wijland}},\ }\href@noop {} {\bibfield
  {journal} {\bibinfo  {journal} {Physical Review Letters}\ }\textbf {\bibinfo
  {volume} {117}},\ \bibinfo {pages} {038103} (\bibinfo {year}
  {2016})}\BibitemShut {NoStop}%
\bibitem [{\citenamefont {Wittmann}\ \emph
  {et~al.}(2017{\natexlab{a}})\citenamefont {Wittmann}, \citenamefont {Maggi},
  \citenamefont {Sharma}, \citenamefont {Scacchi}, \citenamefont {Brader},\
  and\ \citenamefont {Marconi}}]{wittmann2017effective1}%
  \BibitemOpen
  \bibfield  {author} {\bibinfo {author} {\bibfnamefont {R.}~\bibnamefont
  {Wittmann}}, \bibinfo {author} {\bibfnamefont {C.}~\bibnamefont {Maggi}},
  \bibinfo {author} {\bibfnamefont {A.}~\bibnamefont {Sharma}}, \bibinfo
  {author} {\bibfnamefont {A.}~\bibnamefont {Scacchi}}, \bibinfo {author}
  {\bibfnamefont {J.~M.}\ \bibnamefont {Brader}}, \ and\ \bibinfo {author}
  {\bibfnamefont {U.~M.~B.}\ \bibnamefont {Marconi}},\ }\href@noop {}
  {\bibfield  {journal} {\bibinfo  {journal} {arXiv preprint arXiv:1701.09032}\
  } (\bibinfo {year} {2017}{\natexlab{a}})}\BibitemShut {NoStop}%
\bibitem [{\citenamefont {Wittmann}\ \emph
  {et~al.}(2017{\natexlab{b}})\citenamefont {Wittmann}, \citenamefont
  {Marconi}, \citenamefont {Maggi},\ and\ \citenamefont
  {Brader}}]{wittmann2017effective2}%
  \BibitemOpen
  \bibfield  {author} {\bibinfo {author} {\bibfnamefont {R.}~\bibnamefont
  {Wittmann}}, \bibinfo {author} {\bibfnamefont {U.~M.~B.}\ \bibnamefont
  {Marconi}}, \bibinfo {author} {\bibfnamefont {C.}~\bibnamefont {Maggi}}, \
  and\ \bibinfo {author} {\bibfnamefont {J.~M.}\ \bibnamefont {Brader}},\
  }\href@noop {} {\bibfield  {journal} {\bibinfo  {journal} {arXiv preprint
  arXiv:1702.00337}\ } (\bibinfo {year} {2017}{\natexlab{b}})}\BibitemShut
  {NoStop}%
\bibitem [{\citenamefont {Wittkowski}\ \emph {et~al.}(2017)\citenamefont
  {Wittkowski}, \citenamefont {Stenhammar},\ and\ \citenamefont
  {Cates}}]{wittkowski2017nonequilibrium}%
  \BibitemOpen
  \bibfield  {author} {\bibinfo {author} {\bibfnamefont {R.}~\bibnamefont
  {Wittkowski}}, \bibinfo {author} {\bibfnamefont {J.}~\bibnamefont
  {Stenhammar}}, \ and\ \bibinfo {author} {\bibfnamefont {M.~E.}\ \bibnamefont
  {Cates}},\ }\href@noop {} {\bibfield  {journal} {\bibinfo  {journal} {New
  Journal of Physics}\ }\textbf {\bibinfo {volume} {19}},\ \bibinfo {pages}
  {105003} (\bibinfo {year} {2017})}\BibitemShut {NoStop}%
\bibitem [{\citenamefont {Curatolo}\ \emph {et~al.}(2019)\citenamefont
  {Curatolo}, \citenamefont {Zhou}, \citenamefont {Zhao}, \citenamefont {Liu},
  \citenamefont {Daerr}, \citenamefont {Tailleur},\ and\ \citenamefont
  {Huang}}]{curatolo2019cooperative}%
  \BibitemOpen
  \bibfield  {author} {\bibinfo {author} {\bibfnamefont {A.~I.}\ \bibnamefont
  {Curatolo}}, \bibinfo {author} {\bibfnamefont {N.}~\bibnamefont {Zhou}},
  \bibinfo {author} {\bibfnamefont {Y.}~\bibnamefont {Zhao}}, \bibinfo {author}
  {\bibfnamefont {C.}~\bibnamefont {Liu}}, \bibinfo {author} {\bibfnamefont
  {A.}~\bibnamefont {Daerr}}, \bibinfo {author} {\bibfnamefont
  {J.}~\bibnamefont {Tailleur}}, \ and\ \bibinfo {author} {\bibfnamefont
  {J.-D.}\ \bibnamefont {Huang}},\ }\href@noop {} {\bibfield  {journal}
  {\bibinfo  {journal} {BioRxiv}\ ,\ \bibinfo {pages} {798827}} (\bibinfo
  {year} {2019})}\BibitemShut {NoStop}%
\bibitem [{\citenamefont {Kolb}\ and\ \citenamefont
  {Klotsa}(2020)}]{kolb2020active}%
  \BibitemOpen
  \bibfield  {author} {\bibinfo {author} {\bibfnamefont {T.}~\bibnamefont
  {Kolb}}\ and\ \bibinfo {author} {\bibfnamefont {D.}~\bibnamefont {Klotsa}},\
  }\href@noop {} {\bibfield  {journal} {\bibinfo  {journal} {Soft Matter}\
  }\textbf {\bibinfo {volume} {16}},\ \bibinfo {pages} {1967} (\bibinfo {year}
  {2020})}\BibitemShut {NoStop}%
\bibitem [{\citenamefont {Dellago}\ \emph {et~al.}(1998)\citenamefont
  {Dellago}, \citenamefont {Bolhuis}, \citenamefont {Csajka},\ and\
  \citenamefont {Chandler}}]{dellago1998transition}%
  \BibitemOpen
  \bibfield  {author} {\bibinfo {author} {\bibfnamefont {C.}~\bibnamefont
  {Dellago}}, \bibinfo {author} {\bibfnamefont {P.~G.}\ \bibnamefont
  {Bolhuis}}, \bibinfo {author} {\bibfnamefont {F.~S.}\ \bibnamefont {Csajka}},
  \ and\ \bibinfo {author} {\bibfnamefont {D.}~\bibnamefont {Chandler}},\
  }\href@noop {} {\bibfield  {journal} {\bibinfo  {journal} {The Journal of
  chemical physics}\ }\textbf {\bibinfo {volume} {108}},\ \bibinfo {pages}
  {1964} (\bibinfo {year} {1998})}\BibitemShut {NoStop}%
\bibitem [{\citenamefont {Weinan}\ \emph {et~al.}(2002)\citenamefont {Weinan},
  \citenamefont {Ren},\ and\ \citenamefont
  {Vanden-Eijnden}}]{weinan2002string}%
  \BibitemOpen
  \bibfield  {author} {\bibinfo {author} {\bibfnamefont {E.}~\bibnamefont
  {Weinan}}, \bibinfo {author} {\bibfnamefont {W.}~\bibnamefont {Ren}}, \ and\
  \bibinfo {author} {\bibfnamefont {E.}~\bibnamefont {Vanden-Eijnden}},\
  }\href@noop {} {\bibfield  {journal} {\bibinfo  {journal} {Physical Review
  B}\ }\textbf {\bibinfo {volume} {66}},\ \bibinfo {pages} {052301} (\bibinfo
  {year} {2002})}\BibitemShut {NoStop}%
\end{thebibliography}%

\end{document}